\documentclass[twocolumn,trackchanges]{aastex63}

\usepackage{amsmath,amssymb}
\usepackage{enumitem}
\usepackage{hyperref}
\usepackage{float}

\accepted{January 22, 2021}
\submitjournal{AJ}

\usepackage{ulem}

\defcitealias{Kopparapu2014}{K14}
\defcitealias{LUVOIR2019}{L19}
\defcitealias{Bixel2020b}{B20}

\defcitealias{Apai2019a}{Apai et al. (2019a)}
\defcitealias{Apai2019b}{Apai et al. (2019b)}
\defcitealias{Apai2019c}{Apai et al. (2019c)}

\newcommand{\codename}{\texttt{Bioverse}}

\newcommand{\flife}{f_\text{life}}
\newcommand{\Otwo}{O$_2$}

\newcommand{\Tdur}{T_\text{dur}}

\newcommand{\tref}{t_\text{ref}}
\newcommand{\leff}{\lambda_\text{eff}}
\newcommand{\Tref}{T_{*,\text{ref}}}
\newcommand{\Rref}{R_{*,\text{ref}}}
\newcommand{\dref}{d_\text{ref}}

\newcommand{\water}{\text{H$_2$O}}
\newcommand{\fwater}{f^\water}
\newcommand{\fwaterhab}{\fwater_\text{EEC}}
\newcommand{\fwaternonhab}{\fwater_\text{non-EEC}}

\newcommand{\oxygen}{\text{O$_2$}}
\newcommand{\ozone}{\text{O$_3$}}

\newcommand{\fozone}{f_\ozone}
\newcommand{\foxy}{f_{\oxygen}}

\newcommand{\thalf}{t_{1/2}}

\newcommand{\vtheta}{\vec{\theta}}

\newcommand{\Reff}{R_\text{est}}
\newcommand{\aeff}{a_\text{eff}}

\newcommand{\ainner}{a_\text{inner}}
\newcommand{\aouter}{a_\text{outer}}
\newcommand{\deltaa}{\Delta a}

\newcommand{\fhz}{f_\text{HZ}}
\newcommand{\fnothz}{f_\text{non-HZ}}
\newcommand{\fnothzfhz}{(f_{{\text{non-HZ}}}/f_{\text{HZ}})}

\newcommand{\ttotal}{t_\text{total}}

\newcommand{\ee}{\eta_\oplus}

\shorttitle{Bioverse}
\shortauthors{Bixel \& Apai}

\begin{document}

\title{\codename: a simulation framework to assess the statistical power of future biosignature surveys}

\author[0000-0003-2831-1890]{Alex Bixel}
\affiliation{Department of Astronomy/Steward Observatory, The University of Arizona, 933 N. Cherry Avenue, Tucson, AZ 85721, USA}
\affiliation{Earths in Other Solar Systems Team, NASA Nexus for Exoplanet System Science}

\author[0000-0003-3714-5855]{D\'aniel Apai}
\affiliation{Department of Astronomy/Steward Observatory, The University of Arizona, 933 N. Cherry Avenue, Tucson, AZ 85721, USA}
\affiliation{Earths in Other Solar Systems Team, NASA Nexus for Exoplanet System Science}
\affiliation{Lunar and Planetary Laboratory, The University of Arizona, 1629 E. University Blvd, AZ 85721, USA}

\begin{abstract}
Next-generation space observatories will conduct the first systematic surveys of terrestrial exoplanet atmospheres and search for evidence of life beyond Earth. While in-depth observations of the nearest habitable worlds may yield enticing results, there are fundamental questions about planetary habitability and evolution which can only be answered through population-level studies of dozens to hundreds of terrestrial planets. To determine the requirements for next-generation observatories to address these questions, we have developed \codename. \codename\ combines existing knowledge of exoplanet statistics with a survey simulation and hypothesis testing framework to determine whether proposed space-based direct imaging and transit spectroscopy surveys will be capable of detecting various hypothetical statistical relationships between the properties of terrestrial exoplanets. Following a description of the code, we apply \codename\ to determine whether an ambitious direct imaging or transit survey would be able to determine the extent of the circumstellar habitable zone and study the evolution of Earth-like planets. Given recent evidence that Earth-sized habitable zone planets are likely much rarer than previously believed \citep{Pascucci2019}, we find that space missions with large search volumes will be necessary to study the population of terrestrial and habitable worlds. Moving forward, \codename\ provides a methodology for performing trade studies of future observatory concepts to maximize their ability to address population-level questions, including and beyond the specific examples explored here.
\end{abstract}

\keywords{}

\section{Introduction}

The field of exoplanet science stands at an exciting turning point. In the past, most exoplanet surveys aimed only to constrain bulk properties - such as size, period, and mass. Moving forward, several groups are developing concepts for space telescopes which would enable the atmospheric characterization of temperate terrestrial planets. Such concepts include the Large UV/Optical/Infrared Surveyor \citep[LUVOIR,][]{LUVOIR2019}, the Habitable Exoplanet Observatory \citep[HabEx,][]{HabEx2019}, the Origins Space Telescope \citep{Origins2019}, the Nautilus Space Observatory \citep{Apai2019a}, the Large Interferometer for Exoplanets \citep[LIFE,][]{Quanz2018}, and the Mid-Infrared Exoplanet Climate Explorer \citep[MIRECLE,][]{Staguhn2019}. By looking for biosignatures in the atmospheres of temperate Earth-sized planets, these observatories would conduct the first systematic search for life beyond the Solar System.

Next-generation observatories will be able to study some of the closest terrestrial exoplanets in unprecedented detail, but this is only the start of their scientific capability: observatories which can study tens to hundreds of terrestrial planets will allow for the first statistical constraints on the atmospheric, geological, and biological properties of terrestrial planets. Some recent works have explored statistical trends and patterns which may only be evident at the population level. For example, habitable zone models predict patterns in atmospheric CO$_2$ and H$_2$O abundance \citep{Bean2017,Lehmer2020} as well as color and albedo across a range of stellar insolations \citep{Checlair2019}. Venus analogs may have larger apparent radii than their temperate siblings due to their thick, post-runaway greenhouse atmospheres \citep{Turbet2019}. Earth's geological record suggests a possible relationship between the ages and oxygen content of Earth-like planets, assuming their atmospheres evolve similarly to Earth's \citep{Bixel2020b}, and with a large enough sample size of potentially habitable planets, next-generation surveys could place the first constraints on the frequency of life in the universe \citep{Checlair2020}. An understanding of population-level trends will provide context for the interpretation of possible biosignatures on individual worlds and could illuminate their potential false positive (i.e. non-biological) sources \citep{Apai2017, Meadows2018}. To avoid statistical false positive scenarios, efforts must also be made to understand which distinct mechanisms could produce the same apparent trends. For example, an increase in cloud deck altitude with insolation could masquerade as a signature of atmospheric erosion in a sample of transiting exoplanets \citep{Lustig-Yaeger2019}.

Recent research has identified key outstanding questions about terrestrial exoplanets, their planetary systems, and the processes which shape them for which future observatories might provide insights \citep[see the SAG 15 report for an overview of several such questions in the context of direct imaging missions,][]{Apai2017}. For example: what are the processes which shape their atmospheric loss \citep[e.g.,][]{Zahnle2017}? Is the habitable zone wide \citep[e.g.,][]{Kasting1993, Kopparapu2014} or narrow \citep[e.g.,][]{Hart1979}? What is the relationship between planet size and tectonic activity \citep[e.g.,][]{Valencia2007,Dorn2018}? Are habitable planets equally common around stars of different mass and activity levels \citep[e.g.,][]{Shields2016}? Which, if any, of these questions could be answered with a next-generation observatory will depend on its technical design and observing strategy. One important metric is the number of terrestrial habitable zone planets which it could realistically detect, but only a subset of these will be habitable, and even inhabited worlds may vary substantially from Earth in their atmospheric composition and evolutionary history.
Furthermore, deep spectroscopic characterization of individual planets will be time-consuming, so strategic choices must be made as to which planets to characterize and at what wavelengths. For these reasons, analyses based solely on the detection yield predictions of future space mission concepts will provide an optimistic assessment of their statistical power.

To enable meaningful statistical hypotheses which can be tested by future observatories, we have developed \codename. \codename\ estimates the statistical power of next-generation exoplanet surveys to detect and study population-level trends by simulating the underlying planet population, survey limitations, observing biases, and statistical analyses which a future observer would perform on a large set of observations of terrestrial planets. After the following brief description of the code structure, we describe its three main components in Sections \ref{sec:planet_generation} through \ref{sec:hypothesis_testing}. In Sections \ref{sec:example1} and \ref{sec:example2}, we use \codename\ to determine the requirements for next-generation surveys to test the habitable zone concept and study the evolution of Earth-like planets.

\section{Code outline}

\begin{figure*}
    \centering
    \includegraphics[width=0.8\textwidth]{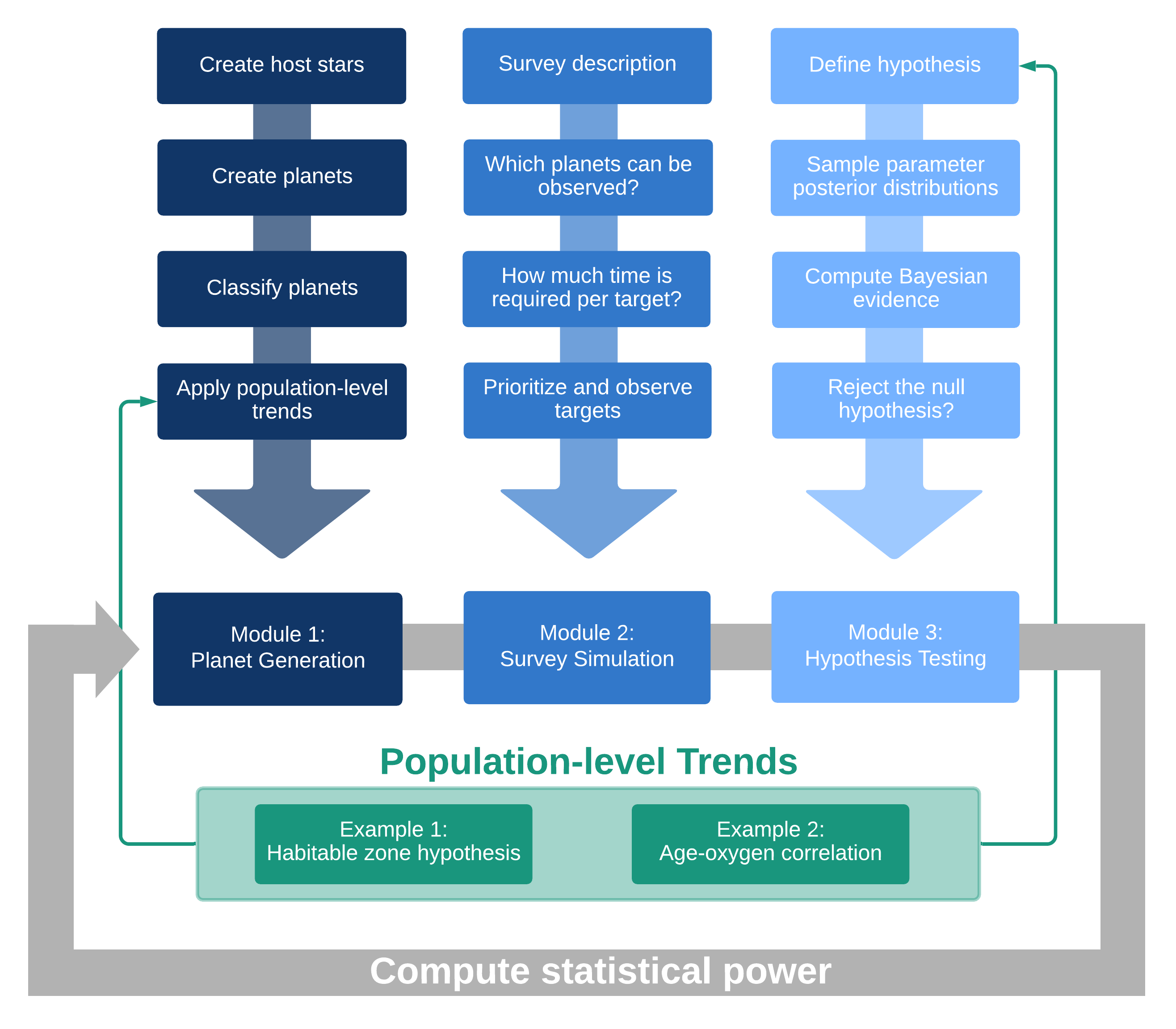}
    \caption{A high-level outline of the \codename\ code. In this paper, we apply \codename\ to assess the detectability of two hypothetical population-level trends (green) with next-generation survey telescopes. These relationships are injected into the simulated planet population by the first module, then tested as statistical hypotheses by the third module.}
    \label{fig:flowchart}
\end{figure*}

\codename\ consists of three components, outlined in Figure \ref{fig:flowchart}. The first component generates planetary systems with bulk properties (e.g., size and period) drawn from statistical distributions, then applies theoretical models or parametric relationships to generate secondary properties of interest (e.g., atmospheric composition). The second component is a survey simulator which conducts observations of the simulated exoplanet population in direct imaging or transit spectroscopy mode. The survey simulator first determines which planets could be characterized within a finite allotted observing time, then generates a simulated data set representative of the telescope and instrument capabilities. The third component is a Bayesian framework which uses simulated datasets to test statistical hypotheses and estimate model parameters. By iterating through these components, we can use \codename\ to determine the statistical power of a proposed observatory to test different hypotheses.

\codename\ is written in Python\footnote{A current version of the code can be found on \href{https://www.github.com/abixel/Bioverse/}{GitHub}, while the version used in this paper is archived on Zenodo \citep{Bioverse}.} and designed for flexibility, so that different statistical assumptions and testable hypotheses can be implemented in the future. The specific set of assumptions which \codename\ is currently based on are listed in Table \ref{tab:assumptions}. Given the large number of parameters involved in \codename, we provide a table of abbreviations and symbols used in the text in Appendix \ref{sec:appendix1}.

\begin{deluxetable*}{p{1.3in}|p{2.7in}|p{2.5in}}
\label{tab:assumptions}
\tablecaption{Summary of statistical assumptions and modeling choices in \codename, with associated references.}
\tablehead{Topic & Assumptions & References}
\startdata
Host star distribution and properties       & (Imaging mode) LUVOIR-A optimized target catalog        & \cite{LUVOIR2019} and C. Stark (private correspondence)                             \\
                                            & (Transit mode) Stellar mass function                          & \cite{Chabrier2003}                   \\
                                            & Main sequence mass-radius-luminosity relations                & \cite{Pecaut2013}                     \\ \hline
Planet occurrence rates                     & SAG13 occurrence rates, with modifications:                   &  \\
                                            & - $\ee \approx 7.5\%$ for G stars (down from $\approx 24\%$)  & \cite{Pascucci2019, Neil2020}         \\
                                            & - More planets around lower-mass stars                        & \cite{Mulders2015a, Mulders2015b}     \\ \hline
Exo-Earth candidates                        & approximately Earth-sized ($0.8 S^{0.25} < R < 1.4 R_\oplus$)  & various (see Section \ref{sec:classification})  \\
                                            & within the circumstellar habitable zone                       & \citetalias{Kopparapu2014}                  \\ \hline
Observatory templates                       & (Imaging mode) 15-meter LUVOIR-A observatory                  & \cite{LUVOIR2019}                     \\
                                            & (Transit mode) 50-meter equivalent Nautilus Space Observatory & \citetalias{Apai2019a}                       \\ \hline
Target prioritization                       & Finite observing time with overheads                          &                                       \\
                                            & Observe in order of required time                    &                                       \\
                                            & Prioritize targets to reduce survey biases                    &                                       \\ \hline
Measurement noise                           & Photon-noise limited observations with characteristic wavelength $\lambda_\text{eff}$                            &                                       \\
                                            & Required exposure time scales with distance, stellar brightness, and signal strength &  \\ \hline
Model comparison                            & Compare alternative to null hypothesis through Bayesian evidence $\mathcal{Z}$                        \\
                                            & Significant evidence when $\Delta(\mathcal{Z}) > 3$                   &                                       \\
                                            & (where applicable) Frequentist comparison tests (e.g., t-tests)    &                                       \\ \hline
\enddata
\end{deluxetable*}

\section{Planet generation}\label{sec:planet_generation}

The first component of \codename\ creates simulated planetary systems around host stars in the solar neighborhood with a period and radius distribution informed by \emph{Kepler} statistics. Other planet properties (such as mass and geometric albedo) are derived from empirical relationships or best-guess prior distributions. Finally, the simulated planet properties reflect the effects of hypothetical population-level trends which could be uncovered by a future survey of terrestrial planets.

\subsection{Stellar properties}
We begin by considering which stars in the solar neighborhood would be targeted by future biosignature surveys. Our strategy for simulating stellar systems is mass-dependent, and therefore depends on the observing technique used by the simulated survey. \codename\ currently considers observations through coronagraphic direct imaging (in ``imaging mode'') and transit spectroscopy (in ``transit mode'').

Direct imaging surveys will primarily target the habitable zones of higher-mass (FGK) stars within the nearest 30 pc, the majority of which have already been cataloged by space-based astrometry missions. Not all of these will be equally valid targets, due to the combined effects of distance and background noise sources, such as zodiacal dust \citep{Stark2019}. Sophisticated simulations for the LUVOIR mission concept \citep{LUVOIR2019} have produced an optimized list of targets whose habitable zones could feasibly be probed for Earth-like planets. In imaging mode, we use an optimized stellar target list for the 15-meter LUVOIR-A concept as the basis for simulating nearby planetary systems (C. Stark, private correspondence).

A survey of transiting habitable zone planets would be most sensitive to planets around low-mass (K and M) stars, as their habitable zone planets are more likely to transit, transit more frequently, and produce a deeper relative transit depth. However, the census of low-mass stars is not complete out to $\sim 100$ pc. Therefore, in transit mode, all stellar masses are randomly drawn from a present-day stellar mass function \citep{Chabrier2003} and distribute them isotropically in space. We do not include any known stars or transiting planets in the transit mode sample; as most nearby transiting planets remain undiscovered, this would have little effect on the overall statistical distribution of host star properties.

In both imaging and transit modes, we relate the stellar mass ($M_*$) to its radius, luminosity, and effective temperature ($R_*$, $L_*$, $T_*$) by interpolating a list of these properties as a function of spectral type \citep{Pecaut2013}. Each star is assigned an age drawn uniformly from 0--10 Gyr, reflecting the (to first order) constant star formation rate in the Milky Way for the past 10 Gyr \citep[e.g.,][]{Snaith2015,Fantin2019,Mor2019}.

\subsection{Period and radius occurrence rates}

\begin{figure}
    \centering
    \includegraphics[width=0.5\textwidth]{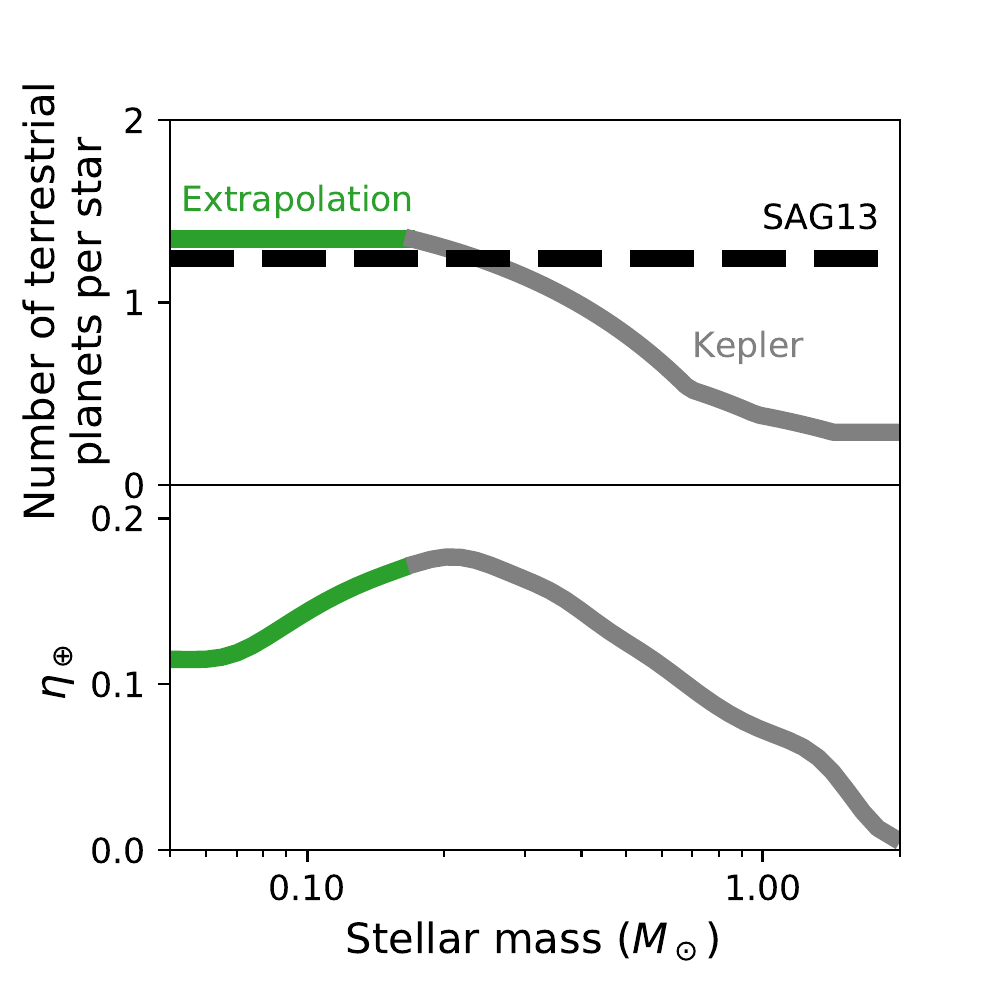}
    \caption{(Top) The assumed number of approximately Earth-sized planets ($0.7 < R < 1.5\,R_\oplus$) with orbital periods shorter than 3 yr per star, as a function of stellar mass. We modify the SAG13 estimate (black) by decreasing the overall planet count by $\sim 3\times$ and increasing the number of planets orbiting \emph{Kepler} low-mass stars, as well as shortening their orbital periods (gray). We conservatively assume the occurrence rates to plateau for ultra-cool dwarfs (green). (Bottom) The corresponding value of $\ee$ using the habitable zone model of \citetalias{Kopparapu2014}.}
    \label{fig:occurrence_rates}
\end{figure}

\emph{Kepler} has provided excellent insights into the frequency of planets as a function of period and size for a wide range of host stars. However, these statistics are only complete to periods $\lesssim 100$ days, and as such do not reach the habitable zone of Sun-like stars. As a result, estimates of $\ee$ (the average number of habitable zone Earth-sized planets per star) have so far been based on extrapolation and are therefore model-dependent. NASA's Exoplanet Program Analysis Group chartered Science Analysis Group 13 (hereafter SAG13) to consolidate the results of several studies of \emph{Kepler} occurrence rates into a single set of estimates for community use,\footnote{see \href{https://exoplanets.nasa.gov/exep/exopag/sag/\#sag13}{this URL} as well as \cite{Kopparapu2018}} resulting in an oft-cited value of $\ee \approx 24\%$ for G stars. Here, and elsewhere in this paper, the value of $\ee$ uses the habitable zone model of \cite{Kopparapu2014} (hereafter \citetalias{Kopparapu2014}; 0.95 -- 1.67 AU for an Earth twin). We use the SAG-13 consensus occurrence rate power laws as the basis for determining the number, radii ($R_p$), periods ($P$), semi-major axes ($a$), and insolations ($S$) of planets in each system. However, the SAG13 metastudy was based largely on studies published before 2017, many of which did not assess planet occurrence as a function of stellar mass. We make the following two modifications to the SAG13 rates to reflect recent work.

First, we unilaterally decrease the number of planets per star by a factor of 3.2, such that $\ee \approx 7.5\%$ for G stars. This is in response to the findings of \cite{Pascucci2019} that Earth-sized planets are more common at shorter orbital periods ($P \lesssim 25$ d) than in the habitable zone, which they ascribe to the effects of photoevaporation. Specifically, they argue that a large fraction of Earth-sized planets on close-in orbits are the evaporated cores of ice giants - planets which maintain their envelopes (and are therefore not Earth-like) if they form in the habitable zone. In another analysis, \cite{Neil2020} find evidence for two distinct populations of rocky planets, and as a result fewer Earth-sized planets in the habitable zone, for which they suggest a similar explanation. The chosen value of 7.5\% is in the mid-range of values estimated by \cite{Pascucci2019} when they exclude the planets most affected by photoevaporation.

Second, we modulate the occurrence rates as a function of spectral type following \cite{Mulders2015a}, who find that rocky planets are more common around lower-mass stars and tend to occupy shorter orbits. Specifically, we gradually increase the number of planets for stars less massive than the Sun and decrease their semi-major axes by interpolating between the scaling factors provided by \cite{Mulders2015a}\footnote{see Table 1 and Figure 4 therein} (normalized to 1 for the typical \emph{Kepler} host star). Later, \cite{Mulders2015b} found evidence that the number of rocky planets around the typical \emph{Kepler} M dwarf (M0 -- M5) was $\sim 3.5\times$ as high as for G dwarfs, so we further increase the number of planets around M dwarfs to reflect this result. Finally, since \emph{Kepler} was not sensitive to late M dwarfs, we assume the number of planets per star to plateau for these stars (which we believe to be a conservative extrapolation given the general trend). We note that more recent studies of M dwarf planet occurrence rates reaffirm the finding that lower-mass stars have more Earth-sized planets, including estimates from \emph{Kepler} data \citep[e.g.,][]{Hardegree-Ullman2019, Hsu2020} and radial velocity detections \citep{Tuomi2019}.

The net impact of these two decisions on the number of Earth-sized planets per star, as a function of stellar mass, is shown in Figure \ref{fig:occurrence_rates}. Our estimate of $\ee$ may seem pessimistic when compared to higher values used in predicting the detection yield of mission concepts \citep[e.g.,][]{HabEx2019, LUVOIR2019, Origins2019, Apai2019a}, but we view it to be a realistic estimate based on the most recent studies available. Our estimate is lower than those of \cite{Bryson2021}, who avoid bias due to photoevaporation by excluding planets at high insolation. However, their resulting sample size is limited, and thus their confidence intervals are broad; indeed, our value of $\ee = 7.5\%$ is within the 95\% confidence interval of some of their estimates.\footnote{see Table 6 therein}

In general, all existing estimates of $\ee$ for G stars - including our own - are based on extrapolation and are therefore uncertain. For example, existing data cannot rule out an increase in terrestrial planet occurrence rates at orbital periods beyond $\sim 100$ d, which would enhance $\ee$. To accommodate this uncertainty, we express our results in Sections \ref{sec:example1} and \ref{sec:example2} in terms of either $\ee$ or the number of planets observed (which is typically linear to $\ee$). As a result, the validity of our results is not tied to any specific value for $\ee$.

\subsection{Habitable zone boundaries} \label{sec:habitable_zone}
The circumstellar habitable zone refers to the theoretical region around a star in which a planet can sustain liquid surface water. Many formulations of the habitable zone exist, but the most commonly cited estimates are based on \cite{Kasting1993} and subsequent papers which expanded on their methodology \citep{Kopparapu2013,Kopparapu2014}. In \codename\, we use the results of \citetalias{Kopparapu2014} to calculate the inner edge ($\ainner$, corresponding to the runaway greenhouse limit) and outer edge ($\aouter$, corresponding to the maximum greenhouse limit) of the habitable zone. To account for the dependence on planetary mass, we interpolate between the three planetary masses modeled therein.

\subsection{Classification} \label{sec:classification}

Following \cite{Kopparapu2018}, we classify planets as ``hot'', ``warm'', or ``cold'' depending on their insolation, and ``rocky'', ``super-Earth'', ``sub-Neptune'', ``sub-Jovian'', or ``Jovian'' depending on their size. Approximately Earth-sized planets within the habitable zone are of particular interest, as these are the most likely planets to have liquid water and habitable surface conditions. Following recent studies of detection yield estimates for direct imaging missions \citep{Kopparapu2018, Stark2019, HabEx2019, LUVOIR2019}, we classify as ``exo-Earth candidates'' (hereafter EECs) any planets with radii $0.8S^{0.25} < R < 1.4$ and orbits within the habitable zone boundaries calculated above. The lower limit on the size of EECs is the theoretical minimum size for which a terrestrial planet can maintain an atmosphere suggested by \cite{Zahnle2017}, while the upper limit reflects the findings of several authors that planets larger than $\sim 1.4 - 1.6 R_\oplus$ tend to resemble mini-Neptunes in composition more than super-Earths \citep[e.g.,][]{Weiss2014, Rogers2015, Fulton2017}.

\subsection{Albedo and contrast ratio}

Imaging measurements will be able to use a planet's brightness as a rough proxy for its size, but its brightness also depends on its geometric albedo, orbital phase, and semi-major axis. The latter two of these can feasibly be constrained by revisiting the system over several months, but it will be difficult to precisely disentangle geometric albedo and planet size. Albedo is highly sensitive to surface and atmospheric composition and will likely be highly variable for directly imaged exoplanets, so estimates of a planet's size based on brightness alone will be highly uncertain \citep{Guimond2018, Bixel2020a, Carrion2020}. To properly represent this source of uncertainty, we assign geometric albedos $(A_g)$ to each planet ranging uniformly from 10 -- 70\% \citep[approximately the range of values encountered at visible wavelengths for solar system planets, e.g.,][]{Madden2018}.

Next, we compute the planet-to-star brightness contrast ratio for each planet, modeling them as Lambertian spheres observed at quadrature phase \citep{Traub2010}:
\begin{equation}
    \zeta = \frac{A_g}{\pi} \left(\frac{R_p}{a}\right)^2
\end{equation}
Note that the determination of a planet's phase from imaging data is also not trivial, requiring multiple follow-up observations to establish the orbit. Nevertheless, such observations will be a likely component of any future imaging survey in order to distinguish temperate planets from their hotter and colder peers \citep{HabEx2019, LUVOIR2019}.

\subsection{Surface gravity and scale height}

To translate planet radii into masses, we use the probabilistic mass-radius relationship derived by \cite{Wolfgang2016}, which separates terrestrial planets and ice giants. Given each planet's mass and surface gravity, we then estimate the atmospheric scale height ($h$), which is important for determining the relative spectroscopic signal due to atmospheric absorption (as described in Section \ref{sec:exptimes}). We assign an atmospheric mean molecular weight $\mu$ to each planet based on its size. For ``sub-Neptune'' planets and larger, we assume H$_2$ dominated atmospheres similar to Neptune or Uranus, with $\mu = 2.5\,m_H$. For ``rocky'' and ``super-Earth'' planets, we calculate the ratio of N$_2$ to CO$_2$ based on their position relative to the habitable zone as follows. For planets within $\ainner$, we assume CO$_2$ dominated atmospheres similar to Venus' ($\mu = 44\,m_H$). Within the habitable zone, we adopt a positive correlation between semi-major axis and CO$_2$ partial pressure, which climate models predict as a result of the carbonate-silicate negative feedback mechanism \citep[e.g.,][]{Bean2017}. Specifically, we follow the correlation derived by \cite{Lehmer2020}\footnote{We adopt the best-fit line in Figure 1 therein}, add N$_2$ as necessary to reach a minimum total pressure of 1 bar, and calculate the mean molecular weight between the two species ($28 < \mu < 44\,m_H$). Finally, for planets beyond $\aouter$, we assume the CO$_2$ to condense, leaving behind a pure N$_2$ atmosphere ($\mu = 28\,m_H$). We set the atmospheric temperature equal to the equilibrium temperature, assuming the Bond albedo to equal the geometric albedo. However, for EECs we assume an Earth-like atmospheric temperature due to greenhouse warming.

\subsection{Inclination and transiting planets}
Planets are assigned inclinations $(i)$ from an isotropic distribution (i.e. a uniform distribution in $\cos(i)$ from $-$1 to 1). From this, and assuming circular orbits, we calculate the impact parameter on the stellar surface:
\begin{equation}
    b = a \cos(i) / R_*
\end{equation}
For transiting planets (with $|b| < 1$) we calculate the transit depth ($\delta = (R_p/R_*)^2$) and duration:
\begin{equation}
    \Tdur =  \frac{R_* P}{\pi a}\sqrt{1-b^2}
\end{equation}

\subsection{Hypothetical population-level trends} \label{sec:model_implementation}
The primary goal of our study is to understand which population-level trends may be detectable with a next-generation exoplanet survey. For example, could such a survey empirically determine the location of the habitable zone based on which planets have $\water$-rich atmospheres (Section \ref{sec:example1}), or study how oxygen evolves over time in the atmospheres of Earth-like planets (Section \ref{sec:example2})?

To enable these inquiries, we apply hypothetical population-level trends to the simulated planet sample which will later be studied by simulated direct imaging and transit surveys. Specifically, we determine which planets have atmospheric water vapor based on their size and semi-major axis (following Equation \ref{eqtn:fwater}), and determine which Earth-like planets have atmospheric oxygen based on their age (following Equation \ref{eqtn:foxygen}). A more detailed description of these assumed trends, and an assessment of their detectability by future biosignature surveys, can be found in Sections \ref{sec:example1} and \ref{sec:example2}.

\section{Survey simulation} \label{sec:survey_simulator}
The second component of \codename\ translates the simulated planet population from the previous section into a data set representing the result of a lengthy characterization effort with a next-generation observatory. There are a few methods by which future observatories could characterize statistically-relevant samples of habitable planets, but in \codename\ we focus on space-based direct imaging and transit spectroscopy. The data sets produced by these next-generation surveys will be inherently biased by the observing approach. Most notably, an imaging survey is most efficient in targeting the habitable zones of nearby FGK stars, while a transit survey is optimized for M stars. Strategic decisions also bias the data set - for example, an imaging survey must dedicate $\sim 4\times$ as much time to study a planet at 2 AU from its star versus an Earth twin, so studying planets near the outer edge of the habitable zone will come at a steep cost.

\subsection{Survey setup}
As our template for a direct imaging survey we use LUVOIR \citep[][hereafter \citetalias{LUVOIR2019}]{LUVOIR2019}, a proposed NASA Flagship-class mission which would use an 8--15 meter segmented mirror and a multi-channel coronagraphic instrument to study terrestrial planets around nearby stars. While the details of the LUVOIR concept have been studied in-depth, our results are based only on its high-level characteristics - specifically, we adopt the 15-meter LUVOIR-A mirror diameter, coronagraphic inner (IWA) and outer (OWA) working angles and noise floor, and the host star catalog used to simulate its detection yield estimates (C. Stark, private correspondence). Our results should be generally applicable to any imaging mission with a similar mirror size and coronagraph.

As our template for a transit survey, we use the Nautilus Space Observatory concept \citepalias{Apai2019a, Apai2019c}, which aims to study transiting exoplanets with the equivalent light-collecting area of a single 50-meter diameter telescope. To achieve this light-collecting power, Nautilus would employ an array of large telescopes with ultralight diffractive-refractive optical elements \citep{Milster2020} (the launch of a single, up to 8.5m diameter telescope has recently been proposed as a NASA Probe-class mission, \citetalias{Apai2019b}). To generate the potential list of transiting planets, we simulate systems to a distance of 150 parsecs, as our simulated surveys tend not to observe targets beyond this distance even when they are available.

Our analyses are based on a 15-meter mirror diameter imaging survey and a 50-meter diameter (equivalent area) transit survey, because among all concepts currently under consideration by the community, these are the ones purporting to offer the largest EEC sample sizes for their respective techniques. It should be noted that a 15-meter imaging survey would also be capable of characterizing nearby transiting planets as a secondary science goal, but we do not model any dual mode surveys here.

\subsection{Which planets can be detected?} \label{sec:yield_estimates}
After simulating a catalog of nearby planetary systems, we discard any planets which cannot be detected by a given mission architecture. In transit mode, we exclude all non-transiting planets. In imaging mode, we exclude all planets whose maximum angular separation is less than the IWA, or whose average angular separation is greater than the OWA, or for which the planet-to-star contrast ratio ($\zeta$) is below the instrument noise floor.

The remaining planets can, in principle, be detected by the survey, but to actually detect most of them will require preliminary observations either using the same telescope architecture or a precursor survey. A dedicated imaging mission would likely be able to detect all of the EECs which it is capable of characterizing during preliminary observations \citep{Stark2019}, but the vast majority of transiting planets within the nearest $\sim 100$ pc remain undiscovered. Most likely, a large-aperture spectroscopic survey of hundreds of transiting planets must be preceded by a space-based all-sky survey, similar to TESS \citep{Ricker2015} or PLATO \citep{Rauer2014} but with sensitivity comparable to \emph{Kepler}. The cost and complexity of such a mission, though considerable, would likely be much less than that of a subsequent characterization effort requiring orders of magnitude greater light-collecting area.

\subsection{Which planets can be characterized?} \label{sec:prioritization}
In-depth spectroscopic characterization is time-consuming, so the number of targets which can be characterized is a function of the total time budget allotted to the characterization effort $(\ttotal)$. Note that $\ttotal$ is not necessarily the same as the total survey lifetime (which might be e.g. 5--10 yr). To determine which planets can be observed within $\ttotal$, we first determine the amount of time required to characterize each planet, including overheads, and prioritize targets based on both their required observing time and their relative importance to the survey's goals.

\subsubsection{Required exposure time} \label{sec:exptimes_ref}
To determine which planets can be characterized within the time budget $\ttotal$, we first determine the amount of exposure time required to spectroscopically characterize a reference planet whose host star properties reflect the typical target for each survey mode. For both observing modes, the reference planet has exactly the same bulk parameters and receives the same incident flux as modern Earth. For direct imaging observations, its star is a nearby solar-type star ($\Tref = 5777$ K, $\Rref = R_\odot$, $\dref = 10$ pc) while for transit observations it is a more distant early M dwarf ($\Tref = 3300$ K, $\Rref = 0.315\,R_\odot$, $\dref = 50$ pc). In the examples to follow, we only consider the detection or non-detection of an absorption feature associated with a species, rather than constraints on the abundance.

We use two general circulation models (GCMs) published by \cite{Komacek2019} to quantify the three-dimensional atmospheric abundance profiles of our reference planets. Both models are water-covered planets around a Sun-like star (imaging mode) or early M dwarf (transit mode) with the same size, mass, and insolation as Earth and 1 bar N$_2$/H$_2$O atmospheres. These models include a treatment of ice and liquid cloud cover, which is an important factor affecting the detectability of molecular features through imaging and transit observations. Notably, because the M dwarf planet is tidally-locked, convection on its dayside is more efficient, leading to strong, high-altitude cloud cover and greater stratospheric $\water$ abundance (T. Komacek, private correspondence). Finally, to enable the analysis in Section \ref{sec:example2}, we inject Earth's modern oxygen abundance (p$\oxygen$ = 20.7\%) into the model atmospheres, reducing the background N$_2$ pressure accordingly.

\begin{figure*}
    \centering
    \includegraphics[width=0.49\textwidth]{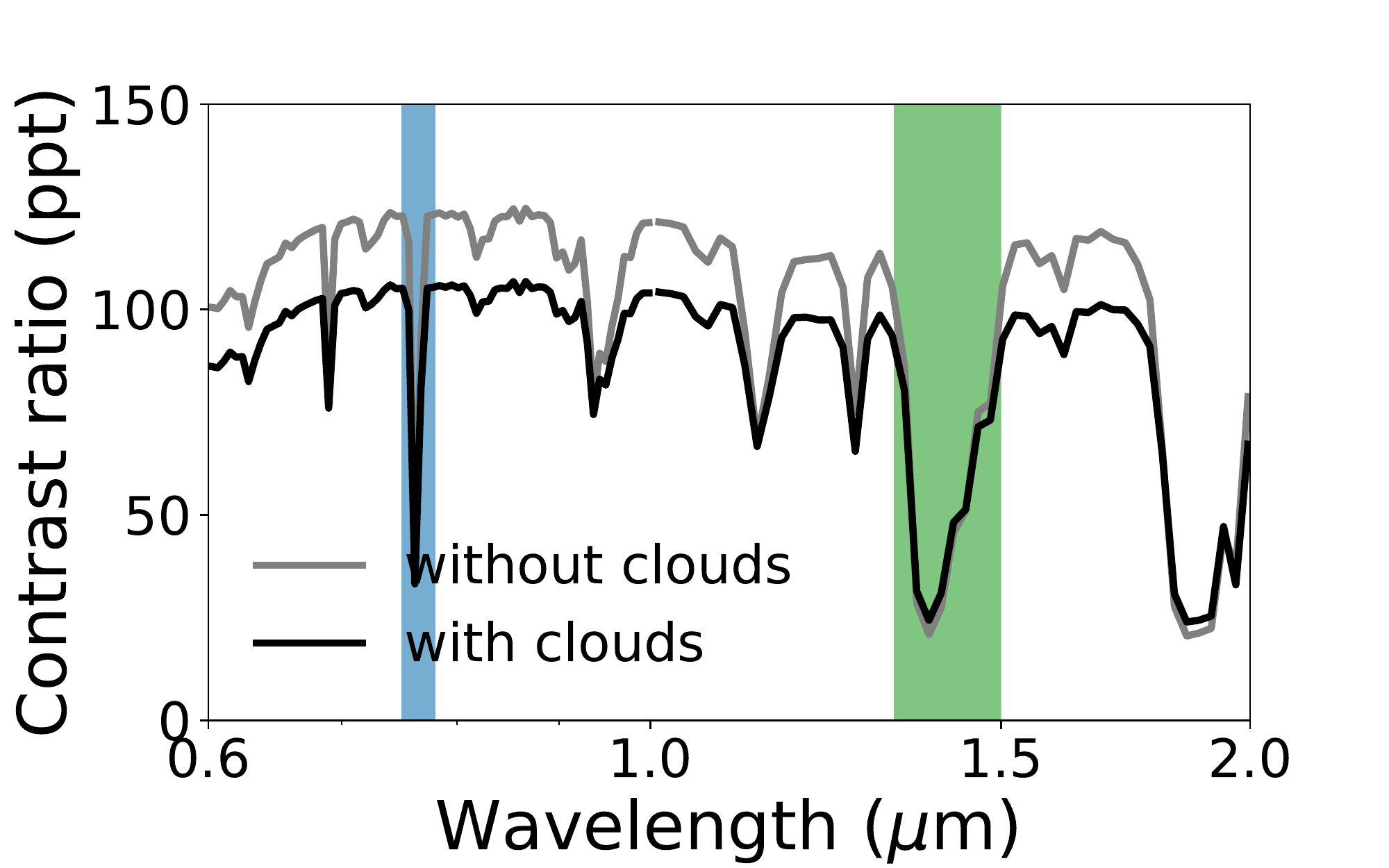}
    \includegraphics[width=0.49\textwidth]{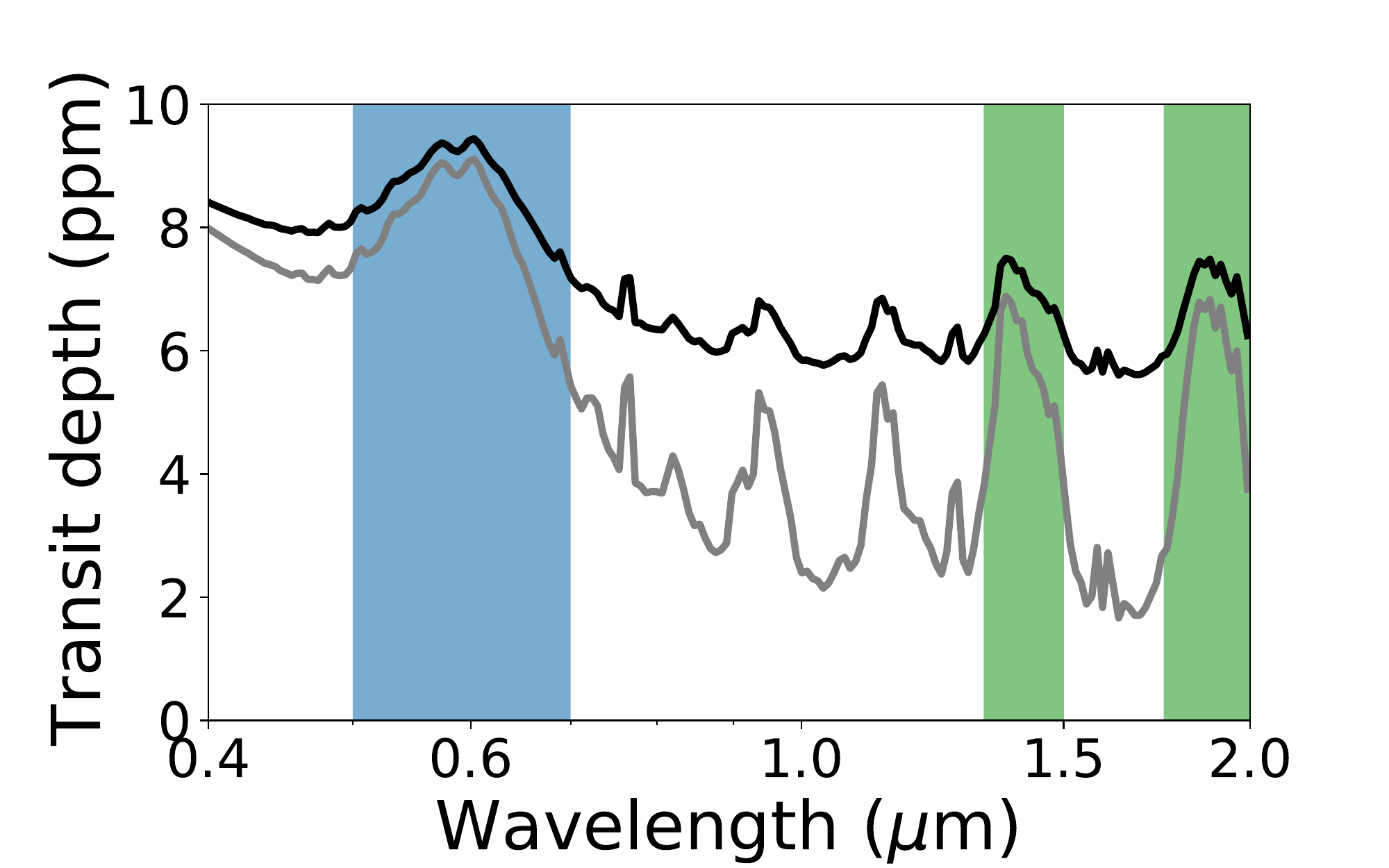}
    \caption{Model spectra for the reference planet in imaging (left; contrast ratio in parts-per-trillion) and transit (right; transit depth in parts-per-million) mode. The spectra are based on GCM models published by \cite{Komacek2019}, who investigate ice and liquid cloud cover on planets as a function of spectral type and tidal locking. We include the effects of clouds to determine our exposure time estimates (black), while clear-sky spectra are shown for reference (gray). Targeted absorption bands include $\water$ (green) and $\oxygen$ or $\ozone$ (blue).}
    \label{fig:clear_cloudy_spectra}
\end{figure*}

To simulate spectra for both models, we use the Planetary Spectrum Generator \citep[hereafter PSG,][]{Villanueva2018}, which accepts three-dimensional atmospheric profiles through its GlobES module\footnote{The PSG configuration files for this study can be found in the code repository}. The directly imaged planet is observed at quadrature phase, while the transiting planet is observed with the night-side facing the observer. Both simulated spectra are shown in Figure \ref{fig:clear_cloudy_spectra} for atmospheres with and without cloud cover. Next, we use PSG to compute noise estimates for each survey architecture as a function of on-target exposure time. In imaging mode, we use the PSG template for the 15-meter LUVOIR-A observatory, including the projected throughput, spectral resolution, raw contrast, and detector noise for the visible and near-infrared imagers, as well as 4.5 zodis of background dust. In transit mode, we simulate observations for a 50-meter diameter aperture with 60\% total throughput, ignoring detector and instrument noise. To determine whether a molecular feature can be detected, we simulate spectra with and without the target molecule and compute the detection signal-to-noise ratio (SNR) across the absorption band in a manner similar to \cite{Lustig-Yaeger2019}\footnote{Equations 4--6 therein}:
\begin{equation}
    \text{SNR} = \sqrt{\sum_i (\Delta y_i/\sigma_{y_i})^2}
\end{equation}
where $\Delta y_i$ is the difference between the two spectra in each spectral bin and $\sigma_{y_i}$ is the measurement uncertainty. Finally, we compute the exposure time required to achieve a SNR = 5 detection of the feature for the reference planet ($\tref$) in each survey mode, then scale this value to determine the exposure time required for each individual planet detected by the survey.

\subsubsection{Exposure time scaling} \label{sec:exptimes}
We define $t_i$ as the amount of exposure time required to spectroscopically characterize a planet at wavelength $\leff$. If we assume that $t_i$ depends primarily on the number of photons collected, then we can estimate it by scaling $\tref$ (as determined using PSG) as follows:
\begin{equation} \label{eqtn:t_scale}
\begin{split}
    \frac{t_i}{t_\text{ref}} &=
    f_i
    \left(\frac{d_i}{\dref}\right)^2
    \left(\frac{R_*}{\Rref}\right)^{-2}
    \left(\frac{B_{*,i}(\leff,T_{*,i})}{B_\odot(\leff,\Tref)}\right)^{-1}
\end{split}
\end{equation}
where $f_i$ summarizes the factors affecting the signal strength unique to each observing mode. In imaging mode, the exposure time is inversely proportional to the planet-to-star contrast ratio (assuming observations at quadrature phase):
\begin{equation} \label{eqtn:t_imaging}
\begin{split}
    f_i^\text{im} = \left(\frac{\zeta_i}{\zeta_\oplus}\right)^{-1}
\end{split}
\end{equation}
In transit mode, the transit depth signal induced by the atmosphere is (to first order) $\Delta \delta \sim (R_p/R_*)^2 (h/R_p)$ \citep{Winn2010} and the required exposure time is inversely proportional to its square:
\begin{equation} \label{eqtn:t_transit}
\begin{split}
    f_i^\text{tr} =
    \left(\frac{h_{i}}{h_\oplus}\right)^{-2}
    \left(\frac{R_{p,i}}{R_\oplus}\right)^{-2}
    \left(\frac{R_{*,i}}{\Rref}\right)^4
\end{split}
\end{equation}
We round up $t_i$ to the next integer multiple of the planet's transit duration, because a transit survey would likely observe complete transits to measure the baseline. Planets are considered to be invalid targets if the total number of required transit observations is greater than either the number of available transits within 10 years or $10^3$.

These scaling relations are meant to capture the main factors affecting the relative exposure time required for each target so as to provide an approximate mapping between the total amount of time dedicated to a survey and the number and distribution of targets it can observe. Ultimately, the primary metric affecting a survey's statistical power is usually the number of EECs characterized, and we translate $\ttotal$ into the number of characterized EECs so the reader can interpret our results as a function of sample size.

\subsubsection{Overheads}
In imaging mode, following \citetalias{LUVOIR2019} we increment each planet's required exposure time by 2 hr to account for slew overheads and overheads associated with wavefront control. These overheads end up being relatively insignificant except for the closest targets. In transit mode, we assume 0.5 hr of slew overheads per observation, plus a total overhead equal to the transit duration for baseline observations before and after each transit event.

\subsubsection{Target prioritization}

Given a limited time budget, it seems reasonable to prioritize observations of planets in order of increasing $t_i$ so as to maximize the number of planets observed. However, prioritizing targets strictly by $t_i$ will lead to a biased sample, especially in the case of transit surveys which are strongly biased towards the detection of close-in planets. To counter-act these biases, we assign a weight $w_i$ to each planet, and calculate its priority as follows:
\begin{equation}\label{eqtn:priority}
p_i = w_i/t_i
\end{equation}
The specific choice of $w_i$ depends on the hypothesis being tested and is discussed in Sections \ref{sec:example1} and \ref{sec:example2}. To create the final simulated data set, we observe targets in order of decreasing $p_i$ until some pre-determined time limit $t_\text{total}$ is reached.

\subsection{Comparison between survey modes}
In the following sections, we use \texttt{Bioverse} to evaluate the statistical potential of direct imaging and transit spectroscopy surveys, but we avoid direct comparisons of their results for the following reasons. First, the technical requirements for and limitations of a direct imaging biosignature survey have been more thoroughly explored due to investments in the LUVOIR and HabEx mission concepts. As a result, our results for the transit survey are likely more optimistic. Second, we do not wish to imply that a survey's statistical power is the only or most important dimension for comparison, as each architecture enables unique capabilities which the other does not.

For the topics discussed here, the primary difference between the two surveys is the number of EECs each can characterize. For the 15-meter imaging survey, this number is 15--20, and is volume- rather than time-limited. This estimate is consistent with that of \citetalias{LUVOIR2019} when adjusted for our updated value of $\ee$ ($\approx 7.5\%$ for G stars). For the 50-meter (equivalent area) transit survey, this number grows with time, with e.g. 60--70 EECs being surveyed for \water\ absorption or $\sim 200$ for \ozone\ absorption given $\ttotal = 2$ yr.

\section{Hypothesis testing} \label{sec:hypothesis_testing}
The third component of \codename\ assesses the information content contained within the simulated data sets from the previous section. This assessment focuses on two primary questions: first, how likely is it that the survey would be able to detect the effects of a statistical trend injected into the simulated planet population (Section \ref{sec:model_implementation})? Second, how precisely could the survey constrain the parameters of that trend? To answer these questions, we rely on a standard Bayesian hypothesis testing approach.\footnote{For a review of Bayesian parameter estimation and model selection in astronomy, we refer the reader to \cite{Trotta2008}.}

\subsection{Null and alternative hypotheses}
Each simulated data set can be thought of as a set of independent variables $x$ and dependent variables $y$. For this section (and the examples to follow), we consider $x$ and $y$ to each represent measurements of a single variable, but this hypothesis testing framework can extend to multivariate measurements as well. The hypothesis $h(\vtheta,x)$ describes the relationship between the $x$ and $y$ in terms of a set of parameters $\vtheta$. The simplest hypothesis is the null hypothesis, in which there is no relationship:
\[
h_\text{null}(\theta, x) = \theta
\]
The null hypothesis is compared to an alternative hypothesis, which proposes a relationship between $x$ and $y$, using a Bayesian parameter estimation and hypothesis testing approach.

\subsection{Likelihood function and prior distribution}
Given a hypothesis $h$, the likelihood function takes on one of two forms. In the case where $y$ is binary (e.g., the detection or non-detection of an atmospheric species), then $h$ is the probability that $y = 1$, and the likelihood function is:
\begin{equation} \label{eqtn:likelihood}
    {\mathcal{L}(y | \vec{\theta})} = \prod_i^N \left[ y_i h(\vtheta, x_i) + (1-y_i) (1-h(\vtheta, x_i))\right]
\end{equation}
Alternatively, if $y$ is a continuous variable measured with normal uncertainty $\sigma_y$, then $h$ predicts the expectation value of $y$, and the likelihood is described by the normal distribution:
\begin{equation} \label{eqtn:likelihood2}
    {\mathcal{L}(y | \vec{\theta})} = \prod_i^N  \frac{1}{\sqrt{2\pi\sigma_{y,i}^2}}  \exp\left(-\frac{(y_i - h(\vtheta, x_i))^2}{2\sigma_{y,i}^2} \right)
\end{equation}
Note that in both example applications of \codename\ to follow, we consider a detection or non-detection as our dependent variable and use the likelihood function defined by Equation \ref{eqtn:likelihood}.

The parameter prior distribution is denoted by $\Pi(\vtheta)$. Given limited prior information about the true values of parameters $\vtheta$, we generally assume uniform or log-uniform distributions spanning the range of plausible values. Further justification for our choice of prior distributions can be found in the examples to follow.

\subsection{Parameter estimation and Bayesian evidence}
For each simulated data set, we sample the posterior distribution of the hypothesis parameters $\vtheta$ using a Markov Chaint Monte Carlo (MCMC) algorithm, implemented by \texttt{emcee} \citep{Foreman-Mackey2013}. This sampling yields measurement constraints of the parameters $\vtheta$. We also use a nested sampling algorithm \citep{Skilling2006}, implemented by \texttt{dynesty} \citep{Speagle2020}, to estimate the Bayesian evidence for the alternative hypothesis:

\begin{equation}
\mathcal{Z} = P(y | h) = \int\mathcal{L}(y | \vtheta) \Pi(\vtheta) d\theta
\end{equation}
To test a hypothesis, we can compare its evidence to that of the null hypothesis, finding evidence to reject the null hypothesis when:
\begin{equation}
    \Delta \ln(\mathcal{Z})= \ln(\mathcal{Z}) - \ln(\mathcal{Z}_\text{null}) > 3
\end{equation}
We choose $\Delta \ln(\mathcal{Z}) > 3$ as our threshold because it corresponds to the common $p < 0.05$ threshold for hypothesis testing with other frequentist tests (e.g., Student's t-test).

It should be noted that \texttt{dynesty} also samples the parameter posterior distributions - so why use \texttt{emcee} to do this separately? In short, nested sampling is optimized to measure $\mathcal{Z}$, while MCMC is optimized to determine the posterior distribution. While \texttt{dynesty} can quickly compute the Bayesian evidence with sufficient accuracy ($\sigma_{\ln(\mathcal{Z})} \lesssim 0.5$), we find it takes significantly longer to converge to the same parameter posterior distributions as \texttt{emcee}. Since we repeat each simulated survey $>100,000$ times, we find this mixed approach to be necessary to achieve both accurate evidence and parameter estimations on a reasonable timescale.

\subsection{Statistical power}
Whether or not an individual simulated survey is able to reject the null hypothesis can often depend on stochastic error; one simulated survey may be able to reject the null hypothesis where another cannot. To summarize our results, we re-run each simulated survey several times under the same set of assumptions and calculate the fraction of survey realizations which achieve a positive result. This metric is also known as the statistical power, and it allows us to assess a survey's statistical potential as a function of both survey parameters (such as total survey duration) and as-yet unknown astrophysical parameters (such as the frequency of habitable planets).\\~\\

This concludes the description of the three primary components of \codename. In the following two sections, we will demonstrate applications of \codename\ to its stated goal of assessing the statistical power of next-generation biosignature surveys.

\section{Example 1: Empirical determination of the habitable zone boundaries} \label{sec:example1}

Models of the habitable zone predict that planets with oceans can only exist within a finite - and perhaps very narrow - range of insolations. An associated prediction is that terrestrial planets in the habitable zone with water-rich atmospheres are the most likely candidates for ocean-bearing worlds. These models will play an important role in the design and target prioritization of next-generation observations; for example, preliminary search strategies for future biosignature surveys often dedicate intensive follow-up to water-bearing habitable zone planets \citep{LUVOIR2019}, while delegating non-habitable zone planets to a lower priority. However, models for the habitable zone have not been tested outside of the solar system, and estimates of its location and width have varied by factors of several over the past few decades.

Could future observatories use data acquired from preliminary observations to test the ``habitable zone hypothesis'' i.e., the hypothesis that planets with water vapor should be more abundant within a narrow and finite range of orbital separations? Further, could these data be used to empirically determine the location and width of the habitable zone? The practical benefit of testing the habitable zone hypothesis would be to make the survey's target prioritization strategy more efficient and to better determine which of its targeted planets are most likely to be habitable. By measuring its boundaries, observers could test the predictions of various habitable zone models, and therefore the physical mechanisms on which they rely. Finally, empirical constraints on the width of the habitable zone will be important for determining the occurrence rate of habitable worlds. Here, we use \codename\ to explore how a survey of atmospheric water vapor could be used to test the habitable zone hypothesis.

\subsection{Model predictions}
Climate models predict a steep decline in water vapor abundance of terrestrial planets outside of the habitable zone. Within the inner edge, an Earth-like planet may undergo a runaway greenhouse as on Venus, leaving behind only a tenuous amount of atmospheric water vapor. Beyond the outer edge, the oceans may freeze, and water vapor would not accumulate except in very low pressure atmospheres which permit its sublimation.

In \codename\ we implement these predictions as follows. We assume that a fraction $\fwaterhab$ of EECs are in fact habitable, meaning they bear surface water and atmospheric water vapor. We also allow a fraction $\fwaternonhab$ of non-EECs to have atmospheric water vapor, serving as a source of noise and ``false positives'' for habitable planets. Then the fraction of planets with atmospheric water vapor can be described as:

\begin{equation} \label{eqtn:fwater}
    \fwater =
\begin{cases}
    \fwaterhab            & \text{if } \ainner < a < \aouter \\
                          & \text{and } 0.8S^{0.25} < R < 1.4 R_\oplus \\
    \fwaternonhab         & \text{if } a < \ainner \text{ or } a > \aouter\\
                          & \text{and } R > 0.8S^{0.25} \\
    0                     & \text{if } R < 0.8S^{0.25}
\end{cases}
\end{equation}

where the habitable zone boundaries and planet size limits are those discussed in Sections \ref{sec:habitable_zone} and \ref{sec:classification}.

\subsection{Simulated survey}

\subsubsection{Measurements}
The imaging and transit surveys perform a set of measurements outlined in Table \ref{tab:measurements} to determine the size and orbital separation of each potential target. In imaging mode, the planet's size is not determinable without prior knowledge of the geometric albedo, so an estimated size ($\Reff$) which assumes Earth-like reflectivity is used as a proxy. In both modes, the orbital separation is converted to the ``effective'' semi-major axis ($\aeff$) for which the planet would receive the same insolation around a Sun-like star.

\begin{figure*}
    \centering
    \includegraphics[width=0.9\textwidth]{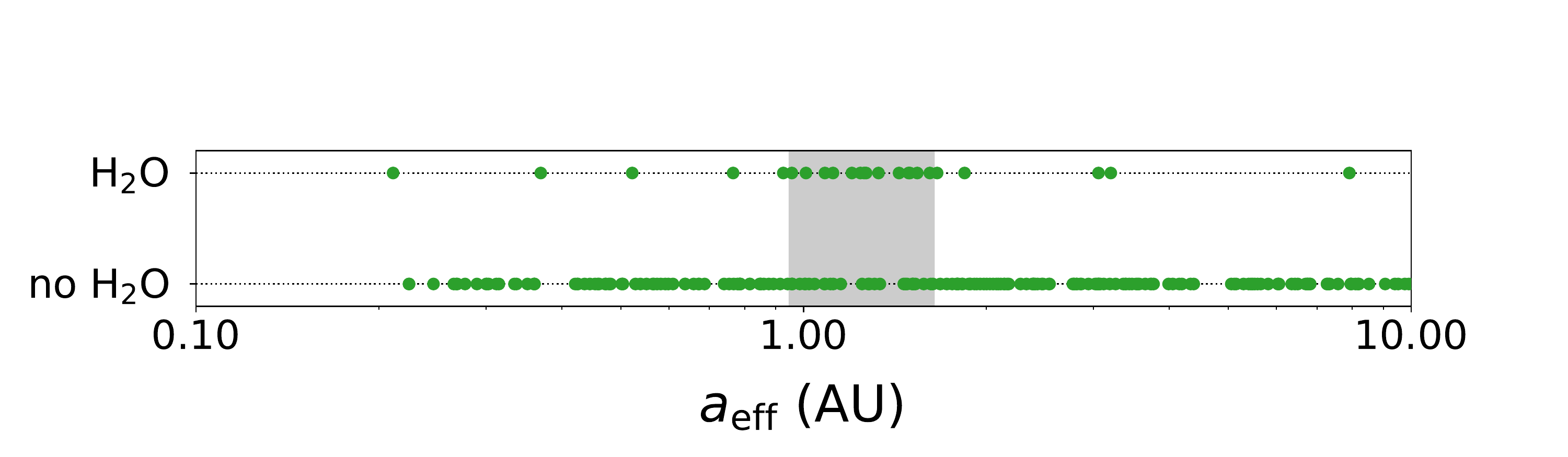}
    \caption{An example of a simulated direct imaging data set for Section \ref{sec:example1}. Planets are probed for the presence of atmospheric water vapor across a broad range of orbital separations. We assume the habitable zone (gray) to be marked by an abundance of water-rich atmospheres. The separation $\aeff = a(L_*/L_\odot)^{-1/2}$ is the solar-equivalent semi-major axis.}
    \label{fig:example1_dataset}
\end{figure*}

These preliminary measurements are used to prioritize targets as discussed in the following section. Those targets of high enough priority are spectroscopically characterized to determine whether their atmospheres contain $\water$. The final output of each simulated survey as a data set consisting of $(\aeff, \water)$, where $\water = \{0, 1\}$ reflects the absence or presence of water absorption features in the planet's spectrum. One example of a simulated data set is shown in Figure \ref{fig:example1_dataset}.

\begin{deluxetable*}{llp{3in}}
\tablecaption{Measurements made by the simulated surveys in Examples 1 and 2. Parameters marked by $\dagger$ are calculated from other measured values.}
\label{tab:measurements}
\tablehead{Parameter & Measurement uncertainty & Description / notes}
\startdata
\sidehead{\textbf{Example 1}}
\textbf{Imaging survey} & & \\
$L_*$           & negligible                & Host star luminosity \\
$\zeta$         & 15\%                      & Planet-to-star contrast \\
$a$             & 10\%                      & Semi-major axis \\
${\aeff}^\dagger$ & $10\%$                  & Solar-equivalent semi-major axis \\
${\Reff}^\dagger$ & $10\%$                  & Estimated radius assuming Earth-like reflectivity \\
H$_2$O          & Detected / not detected   & Presence of 1.4 $\micron$ $\water$ absorption \\
\textbf{Transit survey} & & \\
$M_*$, $R_*$    & 5\%                       & Host star mass and radius \\
$P$             & negligible                & Orbital period \\
$\delta$        & negligible                & Baseline transit depth \\
${\aeff}^\dagger$ & 1.7\%                   & Solar-equivalent semi-major axis \\
$R^\dagger$     & 5\%                       & Planet radius \\
H$_2$O          & Detected / not detected   & Presence of 1.4 $\micron$ and 1.9 $\micron$ $\water$ absorption \\
 & & \\
\sidehead{\textbf{Example 2}}
\textbf{Imaging survey} & & \\
$t_*$                   & 10\%                      & Age (as measured through asteroseismology) \\
$\oxygen$   & Detected / not detected   & Presence of 0.7 $\micron$ $\oxygen$ absorption \\
\textbf{Transit survey} & & \\
$t_*$                   & 30\%                      & Age (model-based estimate) \\
$\ozone$   & Detected / not detected   & Presence of 0.6 $\micron$ $\ozone$ absorption \\
\enddata
\end{deluxetable*}

\subsubsection{Target prioritization}
To test the habitable zone hypothesis we must observe planets spanning a broad range of semi-major axes, but prioritizing targets solely based on required exposure time will bias observations towards close-in planets. Furthermore, planets much smaller or larger than Earth are not likely to be habitable regardless of insolation, and therefore serve as a source of noise. The counter these effects, we weight each target according to its size and orbital separation following Figure \ref{fig:ex1_priority}. We tuned this prioritization based on trial and error to achieve the following goals:
\begin{enumerate}
    \item Prioritize observations of more probable Earth analogs (planets receiving 50--150\% of Earth's incident flux).
    \item Balance observations of widely-separated planets versus close-in planets.
    \item Minimize observations of non-Earth sized planets.
\end{enumerate}
In transit mode, we additionally weight each target by $(a/R_*)$ to negate the bias due to close-in planets being more likely to transit. The resulting distribution of observed planets is also shown in Figure \ref{fig:ex1_priority}.

\begin{figure*}
    \centering
    \textbf{Example 1: Target prioritization and distribution}\\
    \textbf{Imaging survey}\\
    \includegraphics[width=0.6\textwidth]{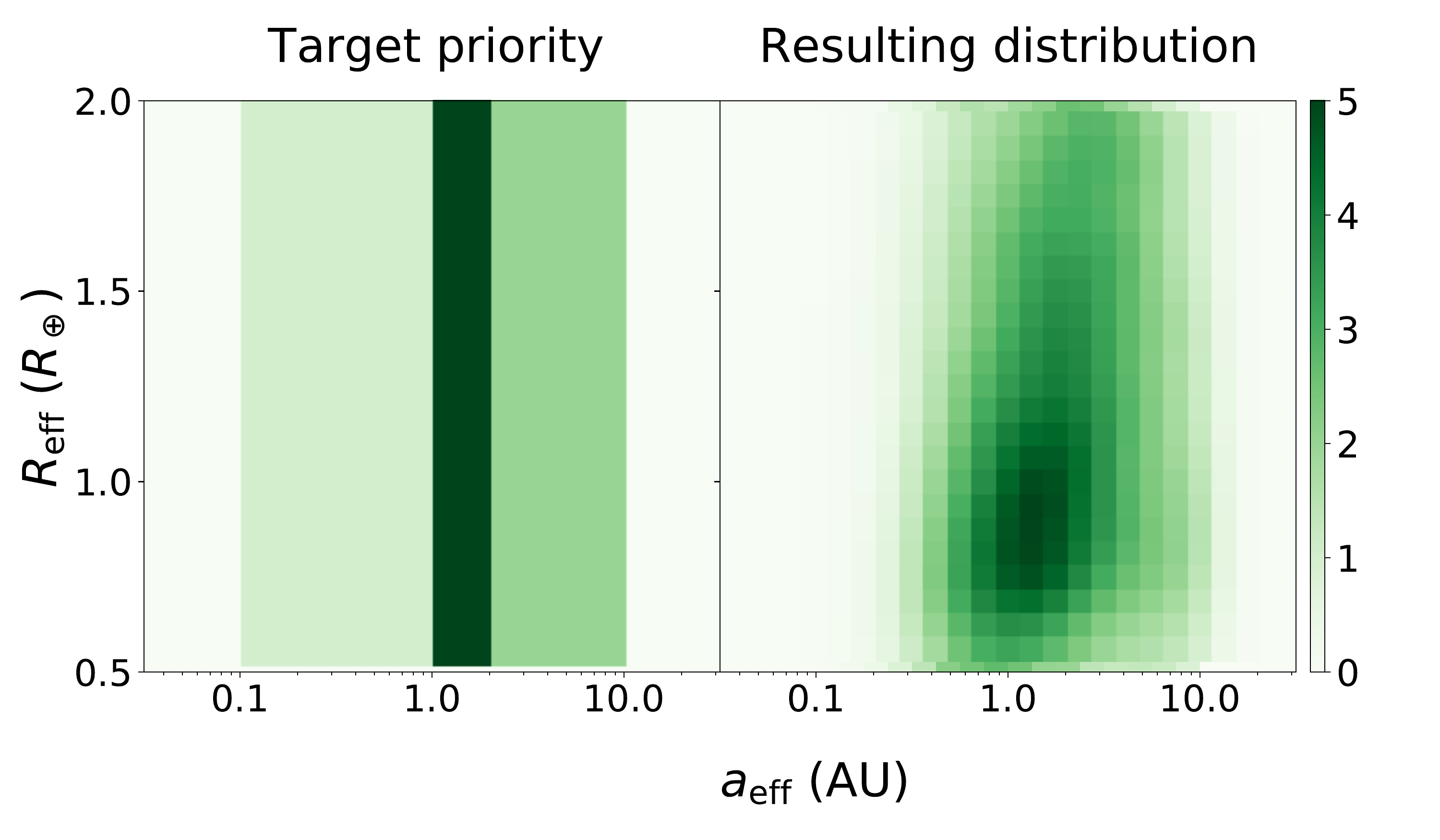}\\
    \textbf{Transit survey}\\
    \includegraphics[width=0.6\textwidth]{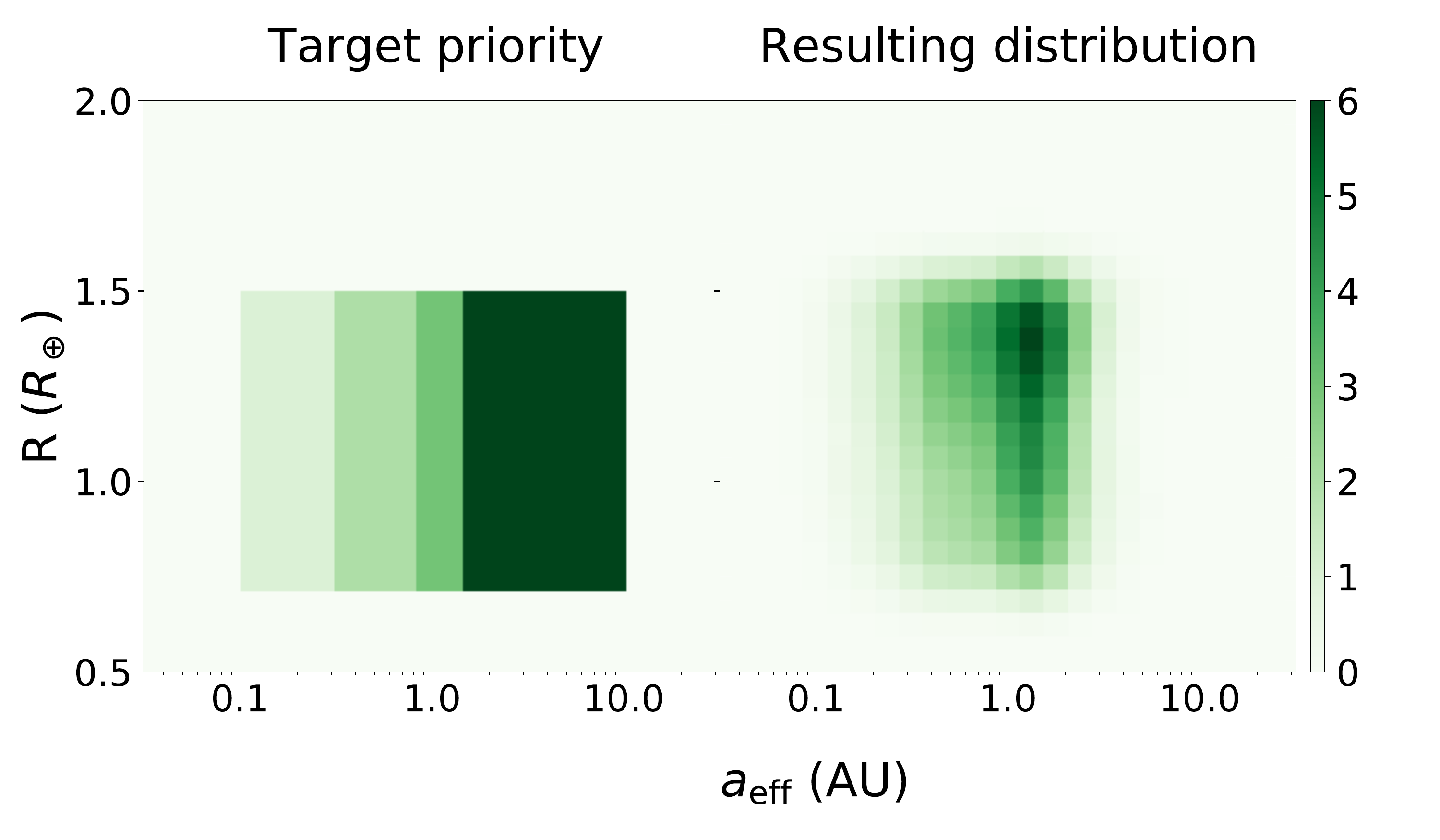}
    \caption{Summary of target prioritization for the simulated imaging (top) and transit (bottom) surveys in Section \ref{sec:example1}. The left panel shows the relative weight assigned to each target as a function of size and orbital separation ($w_i$ in Equation \ref{eqtn:priority}). The right panel shows the resulting relative distribution of targets which can be probed for the presence of water vapor within the survey duration. In the case of the imaging survey, the planet size cannot be directly measured, so the ``estimated'' radius (assuming Earth-like reflectivity) is used as a proxy. In the case of the transit survey, an additional weight is applied to counteract the $R_*/a$ transit probability (not shown above).}
    \label{fig:ex1_priority}
\end{figure*}

\begin{deluxetable*}{lp{2.5in}p{1.5in}}
\tablecaption{Parameter prior distributions for Equations \ref{eqtn:h_HZ} and \ref{eqtn:h_age_oxygen}.\label{tab:priors}}
\tablehead{Parameter & Description & Prior limits (log-uniform distribution)}
\startdata
\sidehead{\textbf{Example 1}}
$\ainner$ & Inner edge of the habitable zone & 0.1 -- 2.0 AU \\
$\deltaa$ & Width of the habitable zone & 0.01 -- 10 AU \\
$\fhz$ & Fraction of habitable zone planets with H$_2$O & 0.001 -- 1 \\
$\fnothzfhz$ & Fraction of non-habitable zone planets with H$_2$O (relative to $\fhz$) & 0.001 -- 1 \\
\sidehead{\textbf{Example 2}}
$\flife$ & Fraction of EECs with life & 0.001 -- 1 \\
$\thalf$ & Oxygenation timescale of inhabited planets & 0.1 to 100 Gyr \\
\enddata
\end{deluxetable*}

\begin{deluxetable*}{llrrrr}
\tablecaption{The predicted signal strengths of $\water$ (Example 1) and $\oxygen$ or $\ozone$ (Example 2) absorption for a representative target of each survey mode, expressed as the peak amplitude of the change in planet-to-star contrast ratio (in parts-per-trillion) or transit depth (in parts-per-million) within the absorption band. The exposure time required for a $5\sigma$ detection is determined using PSG, and scaled for each individual target according to Equation \ref{eqtn:t_scale}. In transit mode, the signals from two bands are combined to achieve the detection of $\water$. In imaging mode, we select the feature which requires the least exposure time to detect. \label{tab:exptimes}}
\tablehead{Survey mode & Feature & Wavelength & Signal strength & Signal strength & Time required (hr)\\ & & & (without clouds) & (with clouds) & (with clouds)}
\startdata
Imaging & $\water$  & 1.4 $\micron$    & 90 ppt  & 55 ppt     & 0.9 \\
        & $\oxygen$ & 0.76 $\micron$    & 90 ppt   & 70 ppt     & 2.6 \\
Transit & $\water$  & 1.4 $\micron$    & 3.5 ppm & 0.5 ppm  & 181 \\
        &           & 1.9 $\micron$    & 5 ppm & 0.7 ppm  &     \\
        & $\ozone$  & 0.6 $\micron$    & 4 ppm & 2 ppm   & 74  \\
\enddata
\end{deluxetable*}

\subsubsection{Time budget} \label{sec:ex1_budget}

Following the procedure in Section \ref{sec:exptimes_ref}, we use PSG to determine the exposure time required for a $5\sigma$ detection of water vapor absorption through its near-infrared absorption bands. In transit mode, we combine the SNR from the 1.4 and 1.9 $\micron$ features. In imaging mode, we only target the 1.4 $\micron$ band, as LUVOIR will be unable to observe the full near-infrared spectrum simultaneously, and the 1.9 $\micron$ band is harder to observe due primarily to lower stellar flux.

In imaging mode, we find $\tref = 0.9$ hr are necessary to probe the reference planet for water vapor, while in transit mode we require $\tref$ = 181 hr  of in-transit exposure time; the details of these calculations are shown in Table \ref{tab:exptimes}. Next, we scale $\tref$ according to Equations \ref{eqtn:t_scale} through \ref{eqtn:t_transit} to determine the amount of exposure time required to characterize each planet, then add observing overheads. We weight the targets according to Figure \ref{fig:ex1_priority}, calculate each target's priority following Equation \ref{eqtn:priority}, then finally observe by order of decreasing probability until the total time budget $\ttotal$ is reached.

The average number of EECs observed by each survey is displayed in Figures \ref{fig:example1_results_imaging}a and \ref{fig:example1_results_transit}a as a function of either $\ee$ or $\ttotal$. While we use the number of EECs observed as our primary metric of sample size, note that most observed targets are non-habitable. Since the imaging survey yield quickly becomes volume-limited, we investigate the impact of varying $\ee$ for a fixed time budget $\ttotal = 120$ d (which is sufficient to characterize $>90\%$ of detectable EECs). For the transit survey, we fix $\ee = 7.5\%$ and investigate the impact of varying $\ttotal$.

\subsection{Habitable zone hypothesis}
Now, let us approach the simulated data from the view of an observer who no prior knowledge of Equation \ref{eqtn:fwater} using the Bayesian hypothesis testing framework outlined in Section \ref{sec:hypothesis_testing}. The habitable zone hypothesis states that planets within the habitable zone are more likely to have water vapor than those outside of it:
\begin{equation} \label{eqtn:h_HZ}
    h^\text{HZ}(\aeff)=
\begin{cases}
    \fhz \quad \text{if } \ainner < \aeff < \ainner + \deltaa \\
    \fhz \fnothzfhz \quad \text{otherwise}
\end{cases}
\end{equation}
This is a four parameter model with $\vec{\theta} = [\ainner, \deltaa, \fhz, \fnothzfhz]$. The choice of parameters was driven by two factors: first, the width of the habitable zone ($\deltaa$) is relevant for testing ``rare Earth'' models in which the habitable zone is very narrow. Second, we can use simple log-uniform prior distributions for these parameters without having to filter out parameter combinations which violate the assumptions of the habitable zone hypothesis (e.g., $\fnothz > \fhz$).

\subsection{Prior assumptions}
\citetalias{Kopparapu2014}'s model for the habitable zone, which we implement in the simulated planet population, assumes a carbon-silicate feedback cycle which enhances CO$_2$ concentrations for planets further from their host stars, and spans 0.95--1.67 AU for the Sun. While this estimate has strong heritage \citep{Kasting1993, Kopparapu2013}, it has also been preceded and succeeded by more conservative or generous estimates, which we use to set the prior distribution of values considered for $\ainner$ and $\deltaa$.

Estimates of the inner edge range as far inward as 0.38 AU \citep[for highly-reflective desert worlds with a minimal greenhouse effect,][]{Zsom2013} and we allow that the inner edge could be as far out as 2 AU, in which case Earth would be an unusually cool outlier. Estimates of the habitable zone width have varied as well; a classic estimate by \cite{Hart1979} suggests a very narrow habitable zone ($\deltaa < 0.1$ AU), which would imply that Earth-like planets are especially rare. More recent estimates have proposed mechanisms by which the habitable zone could extend as far as 2.4 AU \citep{Ramirez2017} or even 10 AU \citep{Pierrehumbert2011} from the Sun. Given this wide range of estimates for $\ainner$ and $\deltaa$, we assume broad prior distributions for both, shown in Table \ref{tab:priors}.

\begin{figure*}[p]
    \centering
    \textbf{Example 1: Results for 15-meter imaging survey} \\

    \textbf{How many exo-Earth candidates are probed for $\water$?} \\
    \includegraphics[width=0.49\textwidth]{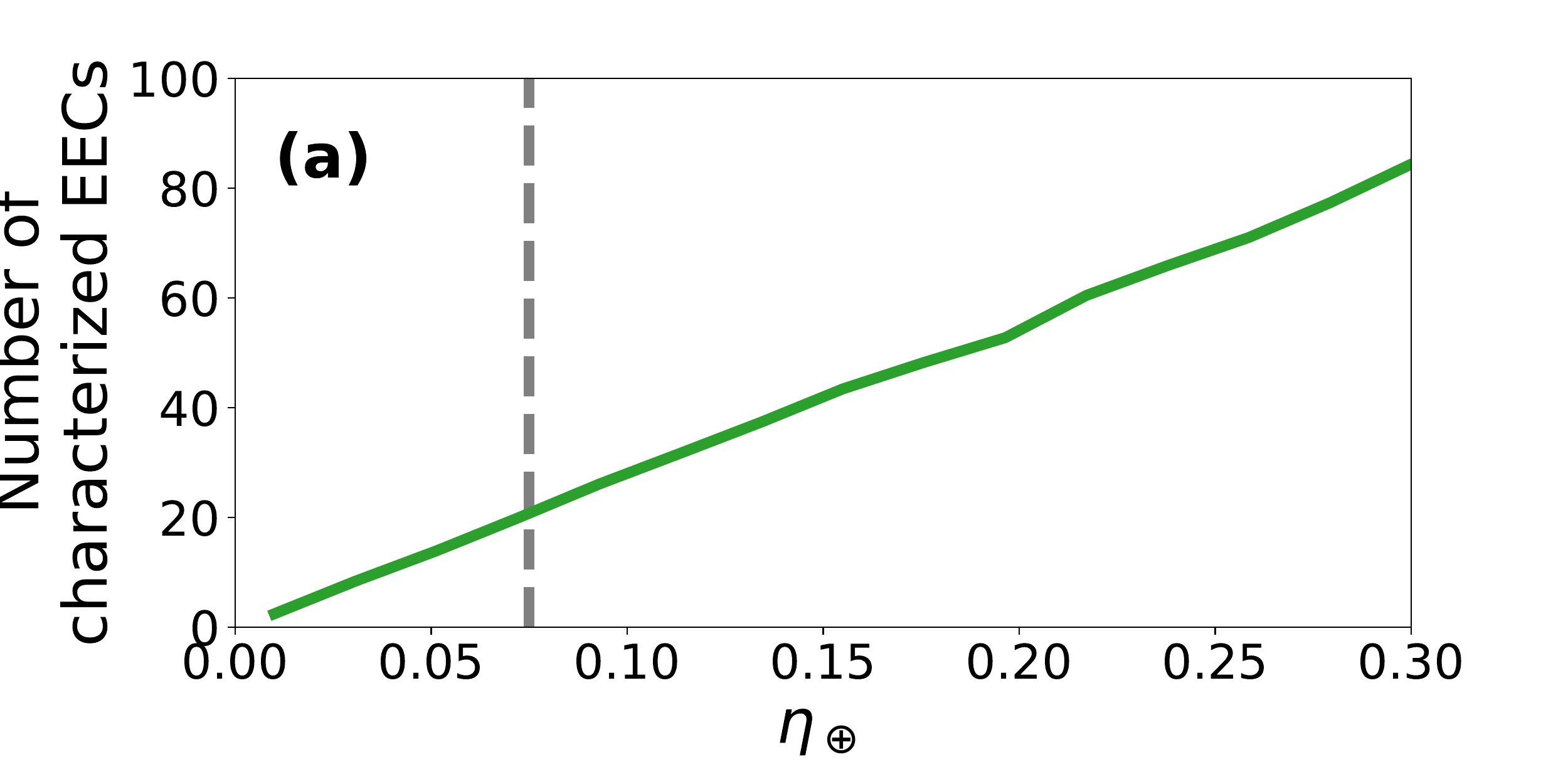}
    \includegraphics[width=0.09\textwidth]{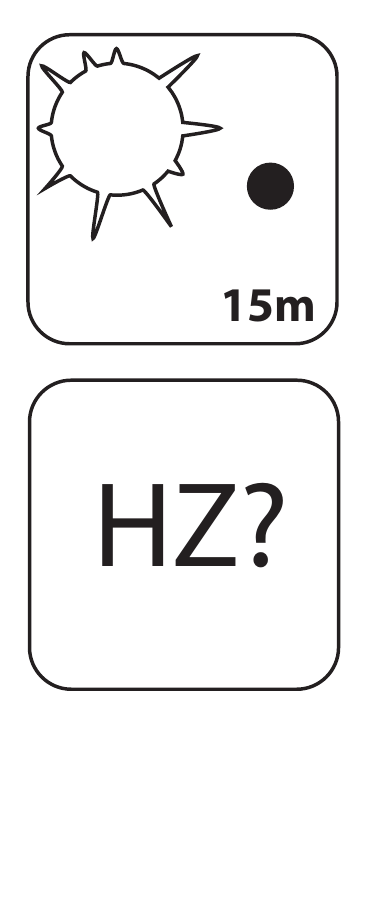}

    \textbf{Can this survey test the habitable zone hypothesis?} \\
    \includegraphics[width=0.49\textwidth]{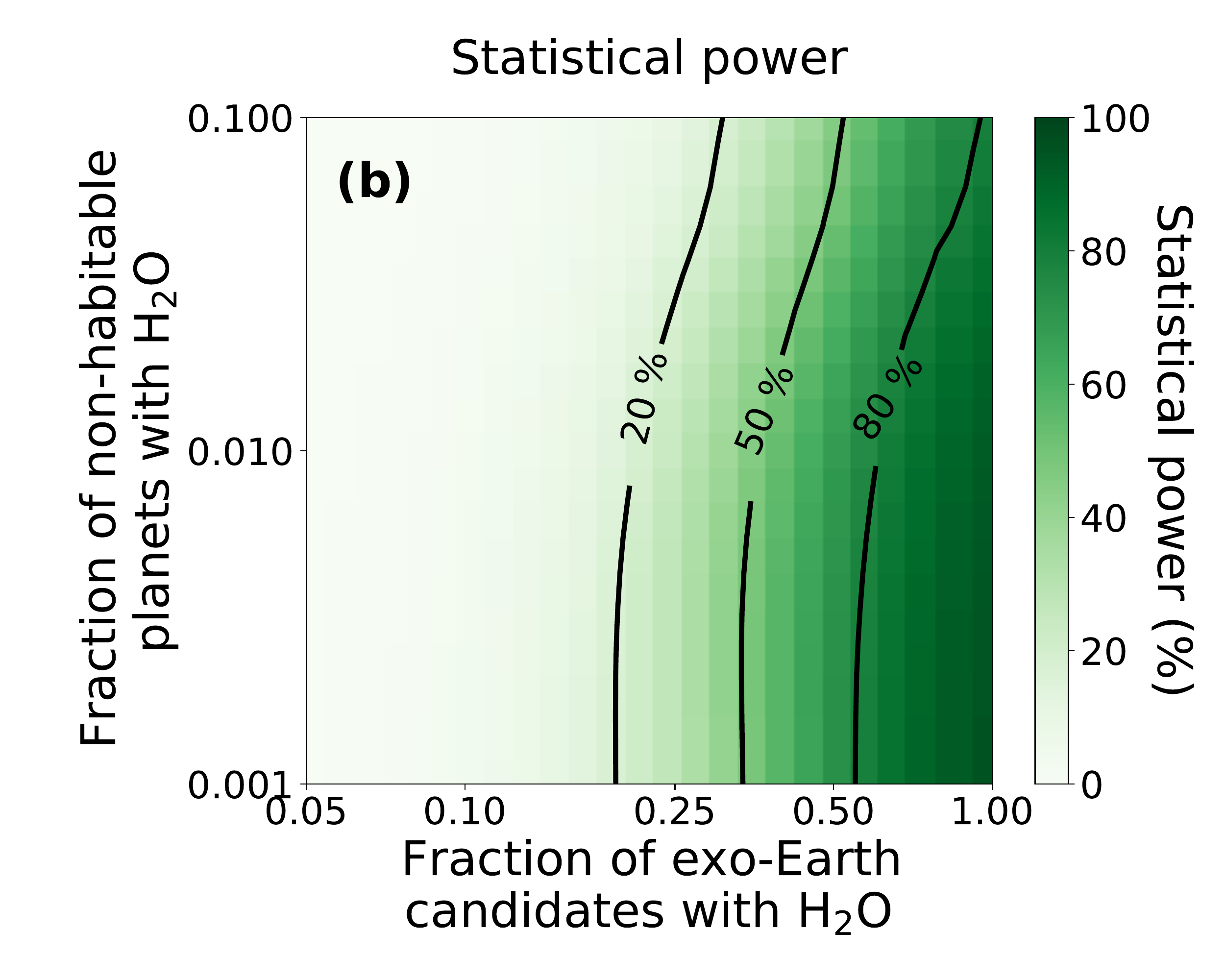}
    \includegraphics[width=0.49\textwidth]{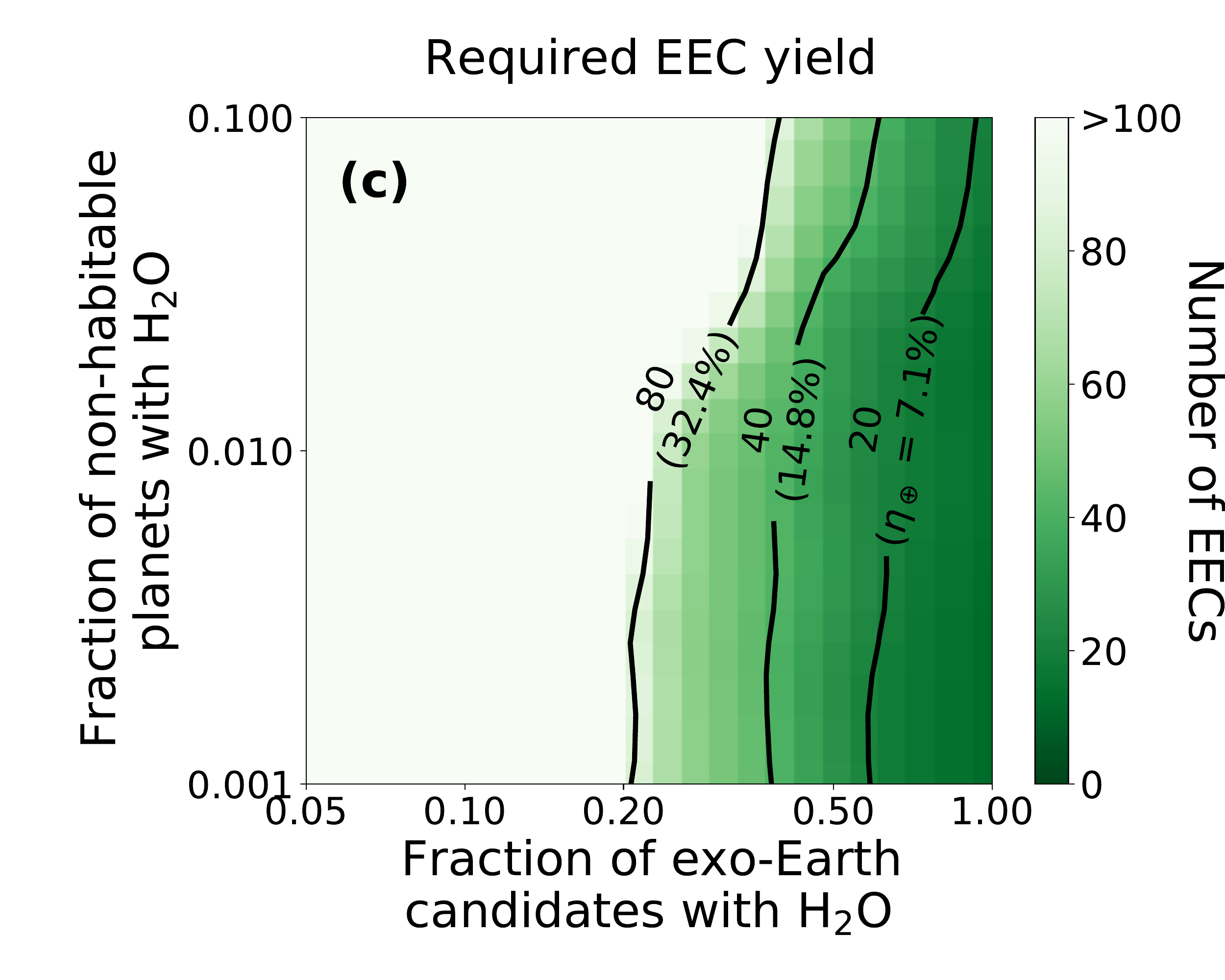}

    \textbf{Can this survey determine the location of the habitable zone?} \\
    \textbf{(results for six random realizations of the survey)}
    \includegraphics[width=0.9\textwidth]{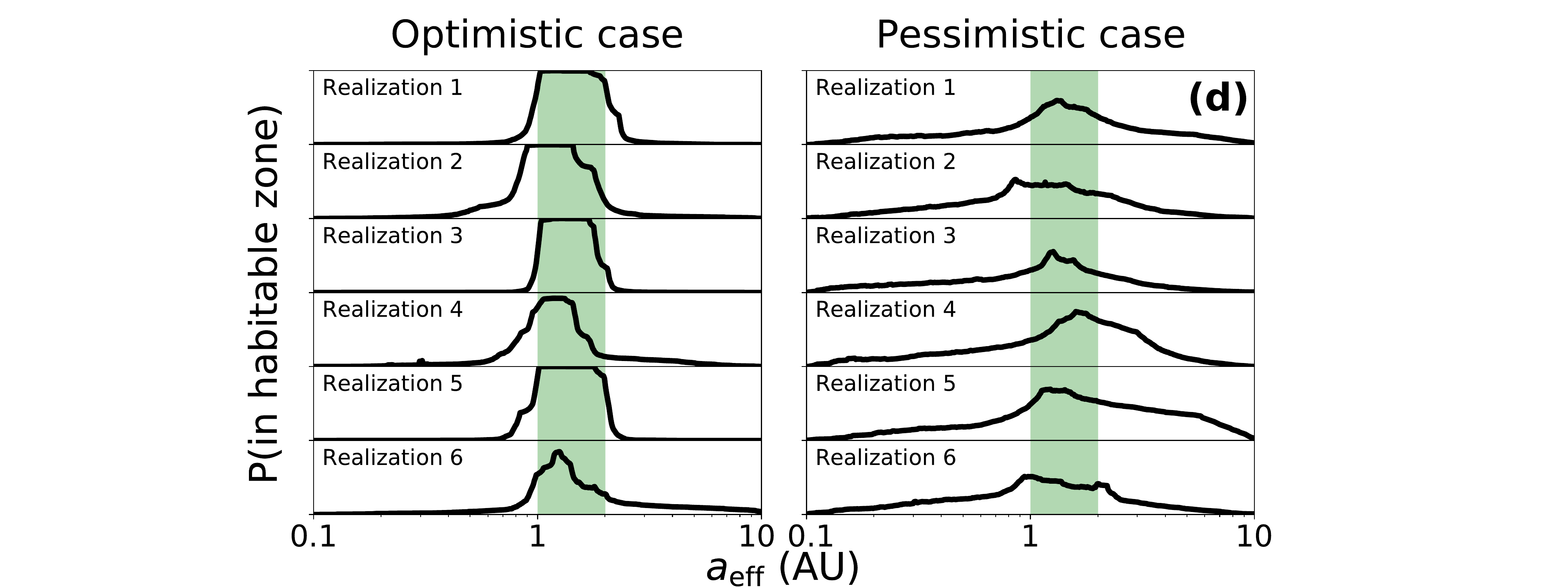}

    \caption{Results for the imaging survey in Section \ref{sec:example1}. (a) The number of EECs observed versus $\ee$ (for G stars), assuming $\ttotal = 120$ d. As our baseline case, we set $\ee = 7.5\%$. (b) The statistical power to test the habitable zone hypothesis as a function of the astrophysical parameters in Equation \ref{eqtn:fwater}. (c) The minimum number of EECs which must be characterized to achieve 80\% statistical power, with the corresponding values of $\ee$. (d) The posterior probability that a planet with effective separation $\aeff$ is in the habitable zone, as estimated by six random realizations of the survey under an optimistic case (80\% of EECs are habitable, left) and pessimistic case (20\% of EECs are habitable, right). The true habitable zone is highlighed in green, and in both cases 1\% of non-habitable planets have $\water$.}
    \label{fig:example1_results_imaging}
\end{figure*}

\begin{figure*}[p]
    \centering
    \textbf{Example 1: Results for 50-meter (equivalent area) transit survey} \\

    \textbf{How many exo-Earth candidates are probed for $\water$?} \\
    \includegraphics[width=0.49\textwidth]{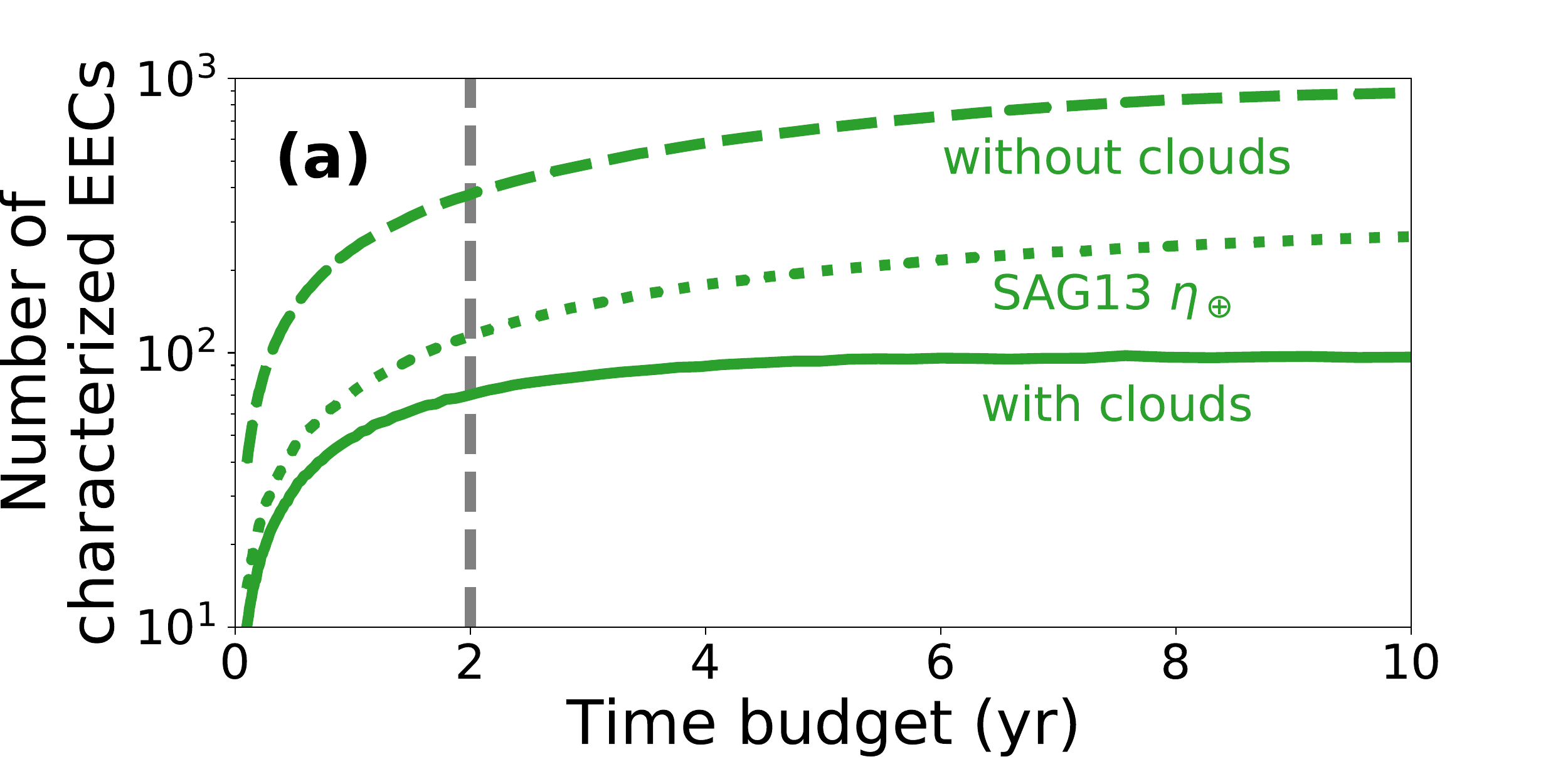}
    \includegraphics[width=0.09\textwidth]{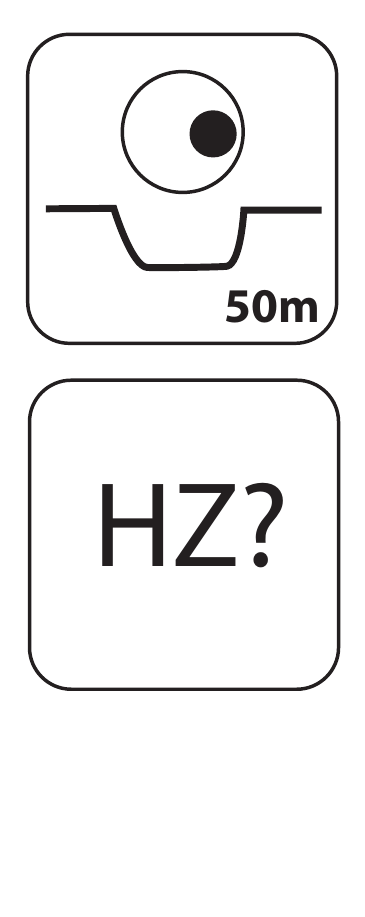}

    \textbf{Can this survey test the habitable zone hypothesis?} \\
    \includegraphics[width=0.49\textwidth]{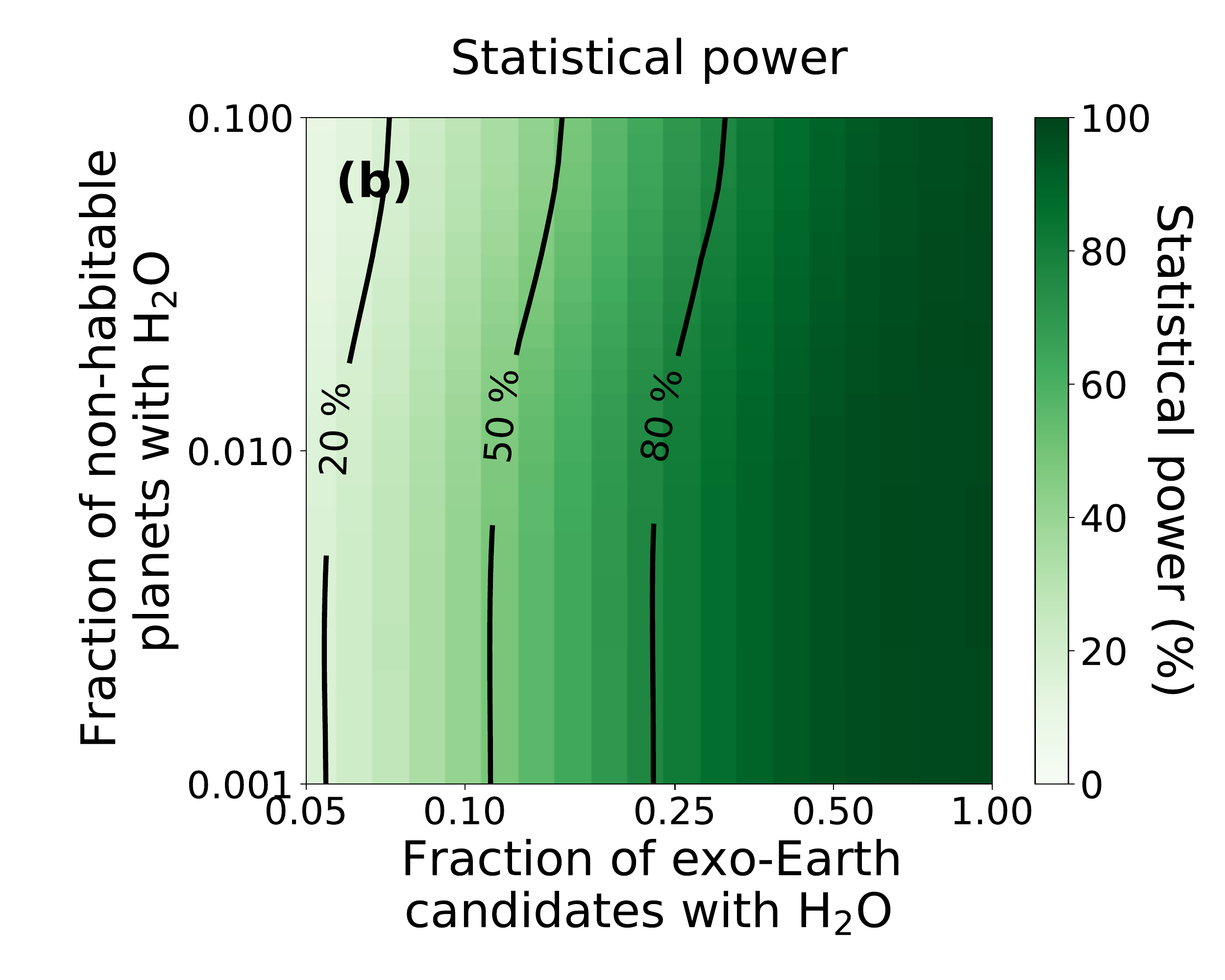}
    \includegraphics[width=0.49\textwidth]{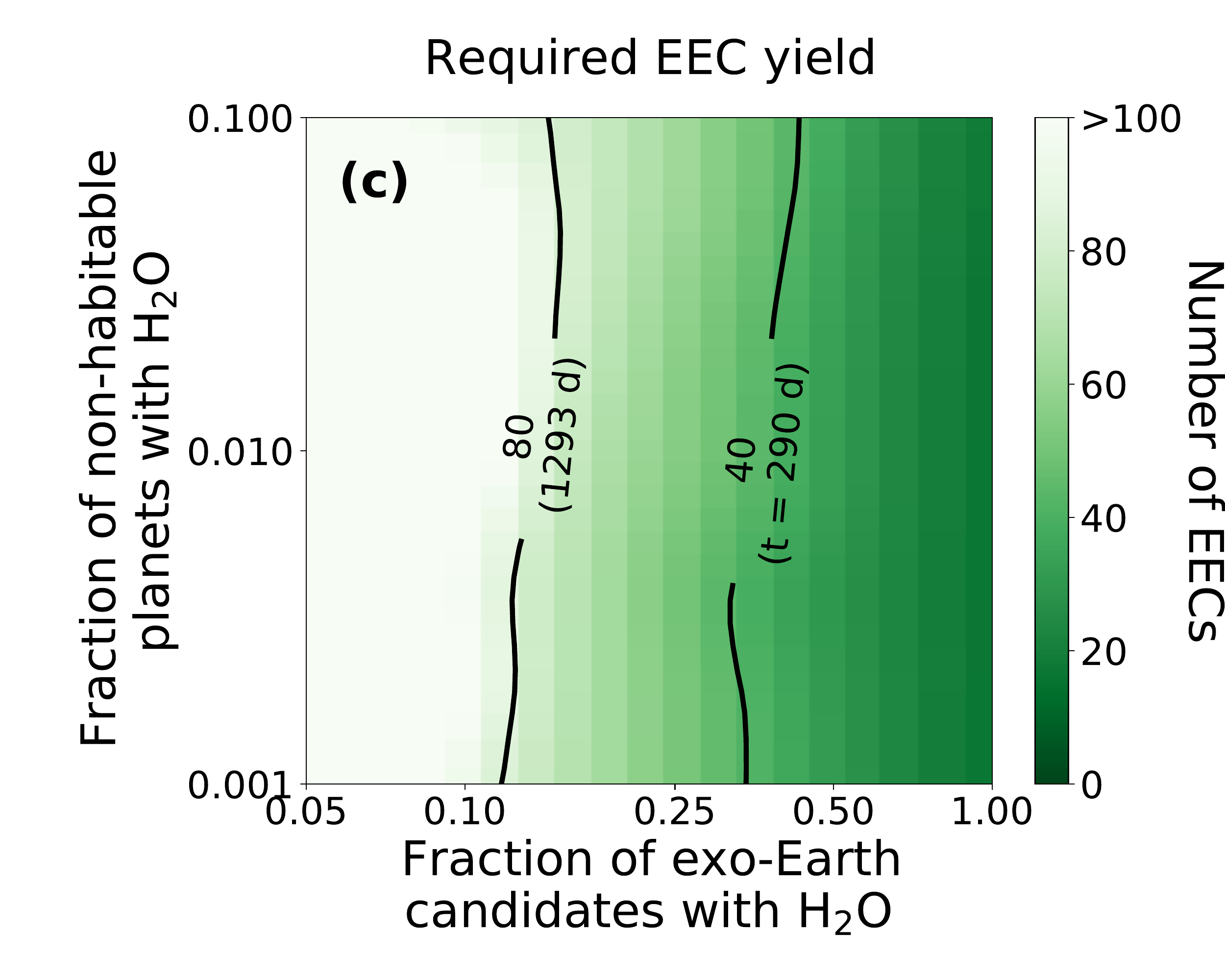}

    \textbf{Can this survey determine the location of the habitable zone?} \\
    \textbf{(results for six random realizations of the survey)}
    \includegraphics[width=0.9\textwidth]{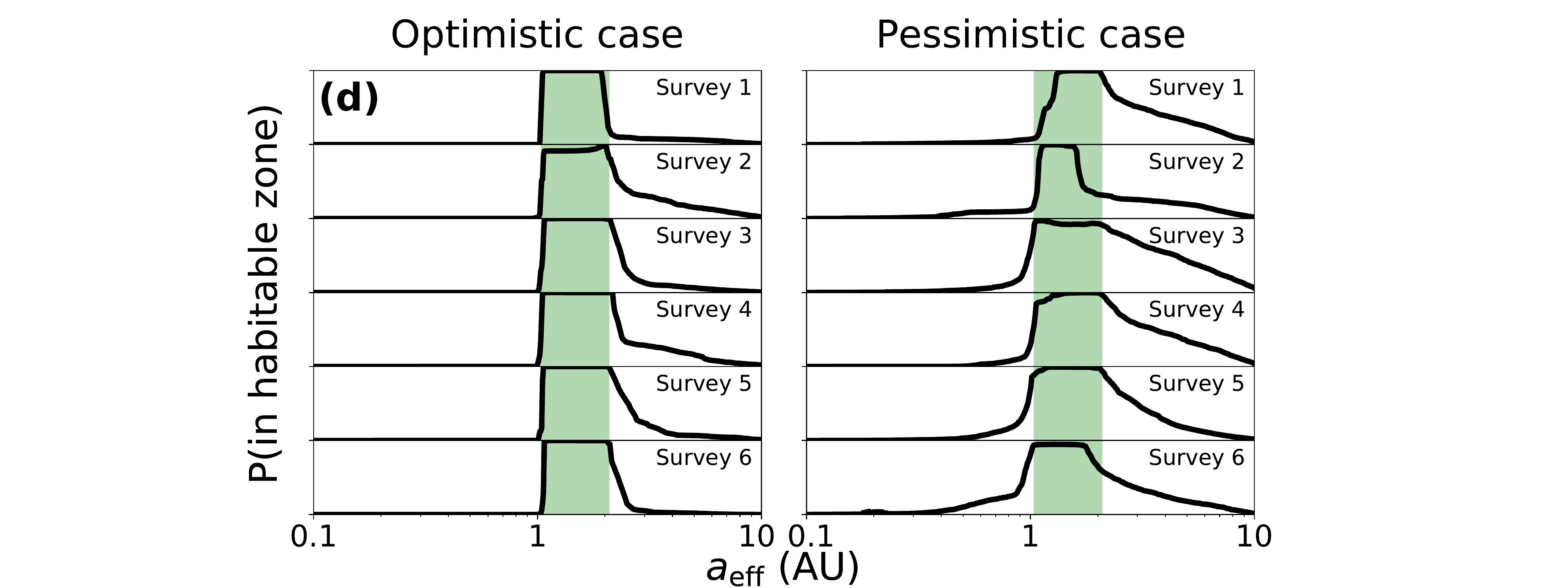}

    \caption{Results for the transit survey in Section \ref{sec:example1}. (a) The number of EECs observed versus the observing time budget, assuming $\ee = 7.5\%$ for G stars and cloudy atmospheres (solid). 4--10$\times$ as many planets could be observed if clouds were neglected (dashed), or 1.5--3$\times$ as many with clouds if assuming the higher SAG13 estimate of $\ee = 24\%$ (dotted). As our baseline case, we set $\ttotal = 2$ yr. (b) The statistical power to test the habitable zone hypothesis as a function of the astrophysical parameters in Equation \ref{eqtn:fwater}. (c) The minimum number of EECs which must be characterized to achieve 80\% statistical power, with the necessary observing time budget $\ttotal$. (d) The posterior probability that a planet with effective separation $\aeff$ is in the habitable zone, as estimated by six random realizations of the survey under an optimistic case (80\% of EECs are habitable, left) and pessimistic case (20\% of EECs are habitable, right). The true habitable zone is highlighted in green, and in both cases 1\% of non-habitable planets have $\water$.
    \label{fig:example1_results_transit}}

\end{figure*}

\subsection{Results}
We repeat the simulated survey and Bayesian analysis $>10,000$ times over a grid of values for the astrophysical parameters in Equation \ref{eqtn:fwater} for both survey architectures. With each simulated survey, we use \texttt{dynesty} to calculate the Bayesian evidence in favor of the habitable zone hypothesis, and \texttt{emcee} to sample the posterior distributions of $\ainner$ and $\deltaa$. The results are summarized by Figure \ref{fig:example1_results_imaging} and \ref{fig:example1_results_transit} for the simulated imaging and transit surveys, respectively.

\subsubsection{Imaging survey}

An ambitious direct imaging survey with a 15-meter telescope could confidently detect the habitable zone with a 3-month long observing campaign provided most EECs are habitable. If habitable planets are less common, however, then more EECs must be observed.  The EEC yield of an imaging survey is typically volume-limited, so higher values of $\ee$ would be required to test this hypothesis for more pessimistic astrophysical parameters. In the best case scenario ($\ee \approx 40\%$), a 15-meter imaging mission could perform the test if 20\% of EECs are habitable, but this value for $\ee$ is likely too optimistic.

If $\sim 80\%$ of EECs are habitable, the imaging survey would be able to measure the location of the habitable zone with sufficient accuracy to exclude some more extreme estimates of its boundaries with reasonable confidence. In particular, it would be able to place a confident lower bound on $\deltaa$, rejecting some ``rare Earth'' models which predict a very narrow habitable zone \citep[e.g.,][]{Hart1979}.

Finally, it should be noted that imaging surveys will have access to planet brightness and color information which could be incorporated into this analysis; for example, albedo and photometric color may vary predictably across the habitable zone \citep{Checlair2019}. Hypotheses which include this information could be tested with better statistical power and parameter constraints than the one examined here.

\subsubsection{Transit survey}

The transit survey can confidently detect the habitable zone even in the case where most EECs are not habitable ($\fwaterhab \approx 25\%$), provided 60--70 EECs can be probed for atmospheric water vapor during a 2-year characterization effort. Furthermore, the full survey duration may not be necessary if most EECs end up to be habitable (in which case a shorter 3--6 month survey would suffice).

The transit survey can precisely measure the inner edge of the habitable zone to within $\pm 0.1$ AU of its true location in most simulated surveys if most EECs are habitable, and can sometimes accomplish this even if most EECs are not habitable. The width (or outer edge) is more difficult to constrain as the transit survey only observes a handful of planets beyond the outer edge ($\sim 10$). This bias has two causes: first, planets outside of the habitable zone are less likely to transit, so they are typically found around more distant stars. Second, colder planets have smaller atmospheric scale heights, and therefore weaker $\water$ absorption features. Both of these effects increase the time required to characterize cold planets, making them low priority targets.

\subsection{Discussion}

\subsubsection{Impact of clouds}

Clouds will have a major impact on the transit survey's ability to test the habitable zone hypothesis, as they dampen the absorption signal due to tropospheric water vapor and therefore increase the number of transit observations required to detect it. As shown in Figure \ref{fig:example1_results_transit}a, this means that a much smaller number of targets can be observed within a fixed time budget, and many of the most distant targets become infeasible to characterize as it would require the combination of decades' worth of transit observations. A possible mitigating strategy would be to expand the observatory's light-collecting area. The Nautilus Space Observatory, on which we base our transit survey results \citep{Apai2019a}, would consist of 35 identically-manufactured unit telescopes. As such, the cost would scale linearly with light-collecting area, and doubling the number of telescopes would reduce by nearly half the number of transit observations required to characterize each planet.

Our cloud assumptions are based on the GCM models of \cite{Komacek2019}, who show that tidally locked planets around M dwarfs have much higher dayside cloud covering fractions than Earth-like planets. If this bears true, it will likely prevent the characterization of such planets through transit spectroscopy by JWST \citep{Fauchez2019, Komacek2020, Suissa2020, Pidhorodetska2020} and possibly even larger observatories. In the pessimistic case, even a 50-meter equivalent area transit survey may be unable to detect atmospheric water vapor for all but a handful of nearby exo-Earths orbiting M dwarfs, so the survey must target more distant K and G dwarfs instead. This will come at the cost of sample size, as we estimate that the increased average distance, less frequent transits, and lower transit depths for habitable zone planets around these stars will outweigh their higher stellar luminosity in terms of observing time cost.

Clouds impact imaging observations as well, although they have little effect on the results presented here. We expect cloud cover to be less prevalent for non-tidally locked planets orbiting Sun-like stars, and highly-reflective clouds at low enough altitudes can have a beneficial effect on imaging observations as they amplify the absorption due to molecules in higher layers. More importantly, the imaging survey modeled here is volume- rather than time-limited, so with or without clouds we find that the survey can probe its entire EEC sample for water absorption within less than three months.

Our exposure time and sample size estimates for the transit survey are based on one possible realization of cloud conditions, but cloud cover may vary greatly across targets and observation epochs. Indeed, in the GCM models we employ, the effect of clouds on transmission spectra is sensitive to orbital period, spectral type, cloud particle size, and many other parameters \citep{Komacek2020}, suggesting that the actual distribution of cloud properties in terrestrial exoplanet atmospheres may be fairly broad. An efficient transit survey could seek to identify planets with clearer atmospheres (e.g. through scattering features in visible light) and prioritize these over cloudier targets, thereby increasing the sample size.

\subsubsection{Effect of non-habitable \water-rich atmospheres}
Naturally, the habitable zone hypothesis is easier to test if more habitable planets are observed, and the number of EEC characterizations required to test it is approximately proportional to the fraction of EECs which are habitable ($\fwaterhab$). However, non-habitable planets are far more common than habitable planets, so if even a small fraction ($\fwaternonhab$) of these have \water, the statistical excess of \water\ in the habitable zone will be muted. In general, we find the statistical power to be unaffected provided that $\fwaternonhab \lesssim 1\%$, but the impact can be considerable if $\fwaternonhab \gtrsim 10\%$. This result seems sensible, as approximately 10\% of the total sample are EECs, so $\fwaternonhab > 10\%$ would imply that \water-rich non-habitable planets are more common than habitable planets.

Our assumption in Equation \ref{eqtn:fwater} is that all EECs with water vapor are habitable, and the fraction of non-EECs with water vapor is mostly independent of insolation. However, if such ``false positives'' exist, their abundance is likely a function of insolation. For example, consider a population of non-habitable planets whose surfaces have been desiccated by a runaway greenhouse effect but which still maintain thick, $\water$-rich atmospheres. Such planets should be clustered near the inner edge of the habitable zone \citep[e.g.,][]{Turbet2019}, appearing as an extension of the habitable planet population to high insolations rather than as a distinct planet population. Even planets defined as EECs may actually be non-habitable (due to differences in initial volatile content, plate tectonics, outgassing rates, etc.) yet still possess water vapor, making them statistically indistinguishable from habitable EECs. Again, the effect of these false positives will likely be negligible provided they are much less common than habitable planets, but indicators of planetary (non-)habitability other than $\water$ may be necessary to filter them out.

\section{Example 2: Evolution of Earth-like Planets} \label{sec:example2}

\begin{figure*}[ht]
    \centering
    \includegraphics[width=\textwidth]{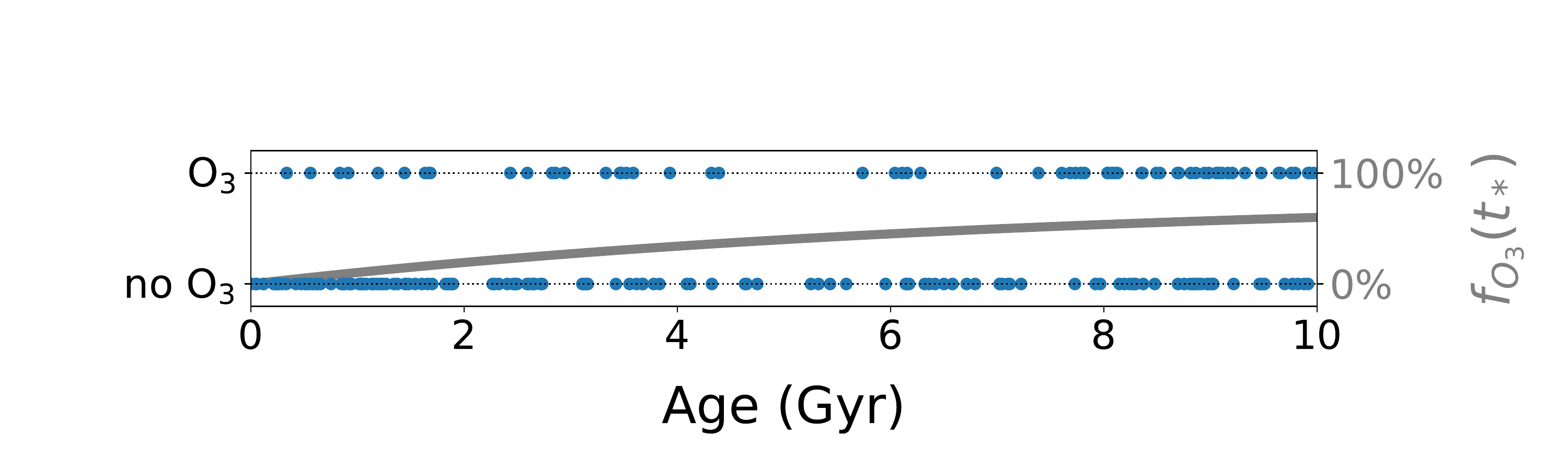}
    \caption{An example of a simulated transit spectroscopy data set for Section \ref{sec:example2}. Earth-sized planets in the habitable zone are probed for the presence of $\ozone$ (a tracer of $\oxygen$), which we assume becomes more common with age as more planets undergo global oxidation events. This ``age-oxygen correlation'' (Equation \ref{eqtn:h_age_oxygen}) is represented by the grey line, in this case where $\flife=80\%$ of observed planets are inhabited and the oxygenation timescale is 5 Gyr. Age estimates are uncertain to $\pm 30\%$.}
    \label{fig:example2_dataset}
\end{figure*}

\begin{figure}
    \centering
    \includegraphics[width=0.5\textwidth]{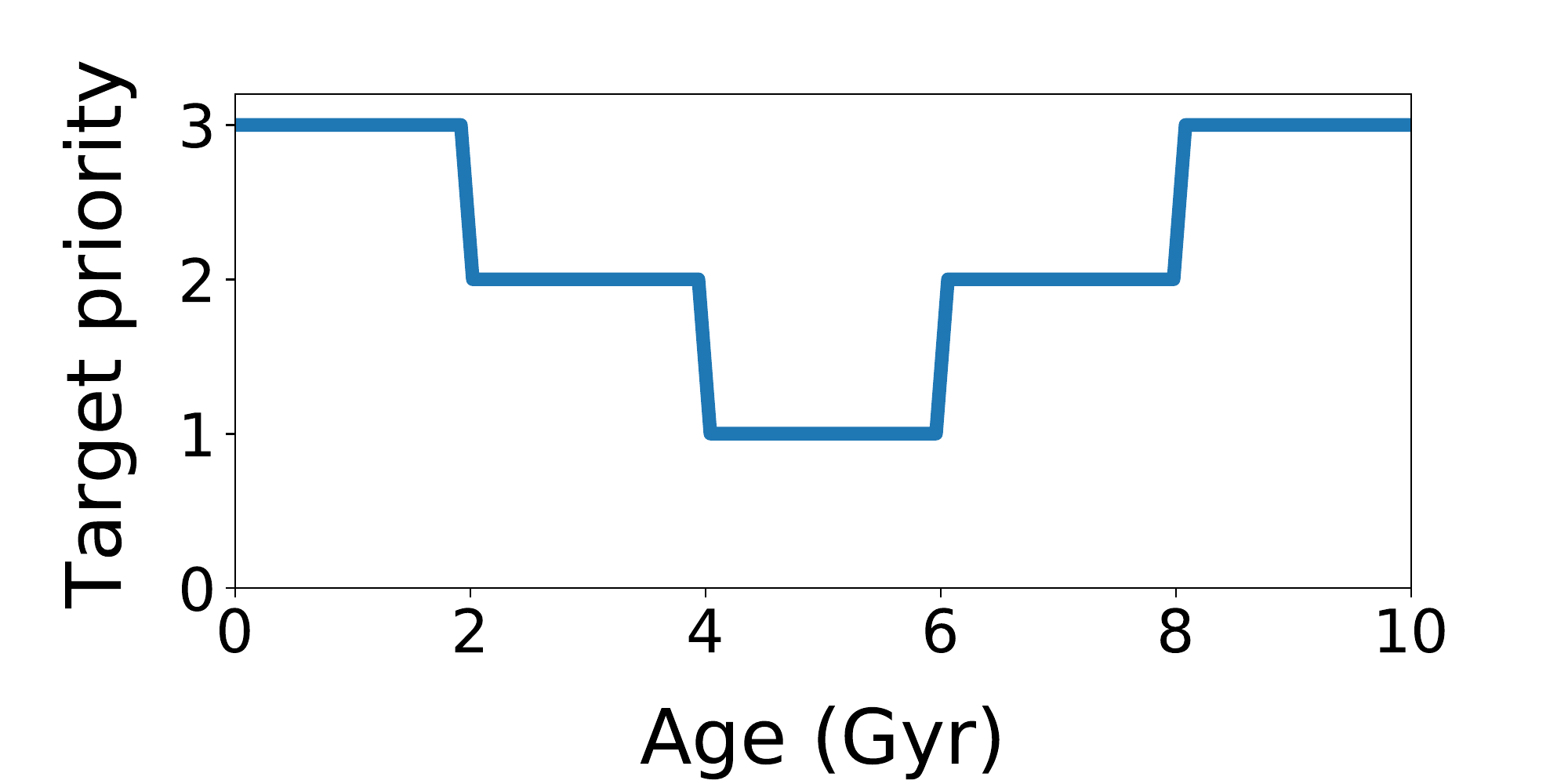}
    \caption{Target prioritization for both surveys in Section \ref{sec:prioritization}, optimized to favor observations of younger and older planets to maximize the detectability of age-dependent trends. This also reflects the age distribution of characterized targets, because the simulated planet sample has a uniform age distribution.}
    \label{fig:ex2_priority}
\end{figure}

By characterizing a sufficiently-large sample of terrestrial worlds, a next-generation observatory could test hypotheses for how they evolve over time. One such hypothesis is that inhabited planets with oxygen-producing life, like Earth, evolve towards greater oxygen content over Gyr timescales due to long-term changes in global redox balance. As we propose in \cite{Bixel2020b} (hereafter \citetalias{Bixel2020b}), the impact on a population level would be a positive ``age-oxygen correlation'', wherein older inhabited planets are more likely to have oxygenated atmospheres.

If inhabited planets do tend to evolve towards greater oxygen content over time, then what is the typical timescale for this evolution? Earth underwent major oxygenation events at 2--2.5 Gyr of age and again at $\sim 4$ Gyr \citep{Lyons2014}, suggesting a $\sim 4$ Gyr ``oxygenation timescale'' \citep{Catling2005}. These two events mark the boundaries between the Archean, Proterozoic, and Phanerozoic eras, and correspond to shifts in Earth's redox balance where the amount of oxygen being produced by life became large enough, and/or the geological sinks for oxygen became diluted enough, that oxygen was allowed to build up in the atmosphere. However, a great diversity of planetary factors might affect redox balance, such as outgassing rates, the stellar radiation profile, biogenic oxygen flux, and the planet's initial reducing matter inventory \citep{Catling2005, Bixel2020b}. As a result, Earth's oxygenation timescale could be unusually fast or slow compared to the overall population of inhabited worlds.

A next-generation biosignature survey could not only detect the proposed age-oxygen correlation, but also measure the typical timescale over which this evolution occurs. This measurement could be used to test models for the geological and biological evolution of Earth-like planets and offer insight into how Earth relates to the rest of that population. Here, we assess the ability of direct imaging and transit surveys to study the oxygenation history of Earth-like planets. This section follows a similar methodology to our previous analysis \citepalias{Bixel2020b}, but expands upon it by incorporating a more thorough assessment of planet occurrence rates, detection sensitivity, and survey strategy, and by studying a broader range of evolutionary timescales.

\subsection{Model predictions}
We assume a fraction $\flife$ of EECs to be inhabited by life - note that this parameter absorbs factors affecting both the planet's habitability and the likelihood of life originating. Over time, simulated inhabited planets transition from anoxic to oxygenated atmospheres at an average rate described by a half life $\thalf$. The resulting fraction of habitable planets which have oxygenated atmospheres as a function of age $t_*$ is:
\begin{equation} \label{eqtn:foxygen}
    \foxy(t_*) = \fozone(t_*) = \flife \left(1 - 0.5^{t_*/\thalf}\right)
\end{equation}
Note that we assume oxygenated atmospheres to have both $\oxygen$ and its photochemical byproduct $\ozone$. We run simulations for $\flife$ ranging from 0--100\% and for $\thalf$ ranging from 500 Myr -- 50 Gyr.

\subsection{Simulated survey}
\subsubsection{Measurements} \label{sec:ex2_measurements}
The measurements performed by each simulated survey are summarized in Table \ref{sec:example2}. First, we measure the age ($t_*$) of every planet's host star with 10\% precision for the imaging survey and 30\% precision for the transit survey. These estimates represent the state of the art for high- and low-mass stars, respectively. For high-mass stars, asteroseismology has yielded highly precise age constraints for \emph{Kepler} targets \citep[e.g.,][]{Creevey2017, Kayhan2019, Lund2019}, and will likely be able to do so for most of the $\mathcal{O}(100)$ stellar targets probed by an imaging mission. For low-mass stars, asteroseismology has not been successful \citep[e.g.,][]{RodriguezLopez2015,Rodriguez2016, Berdinas2017}, and age determination currently relies on a synthesis of model-based estimates. As an example, \cite{Burgasser2017} use a combination of approaches to determine the age of TRAPPIST-1 planetary system \citep{Gillon2017} with $\sim 30\%$ precision.

Next, each planet is observed to constrain the presence of oxygen. For an Earth-like planet, $\oxygen$ can be detected directly through its $0.77\,\micron$ absorption feature or inferred through absorption by stratospheric ozone in the Chappuis (0.40--0.65 $\micron$) or Hartley (0.2--0.3 $\micron$) bands. It should be noted that our calculations assume modern Earth $\oxygen$ and $\ozone$ abundances, an assumption which we revisit in Section \ref{sec:ozone}.

For each survey mode, we determine which of these three features would be easiest to observe across the full range of detected EECs. In imaging mode we observe $\oxygen$-A absorption; while the Hartley band may be easier to detect for a solar-type star, it becomes more expensive to observe for lower-mass stars, and the Chappuis band signal is too shallow. Ultimately this consideration is unimportant for the volume-limited imaging survey, and it is likely that all three features will be searched for in the atmospheres of all detected EECs. In transit mode we observe the Chappuis band, as its signal is strong in transit observations. The Hartley band is inaccessible for the vast majority of (predominantly M dwarf) transit survey targets, and the $\oxygen$-A feature is too shallow and narrow to detect for distant targets.

In total, the simulated surveys produce measurements of $(t_*, \oxygen)$ for each observed EEC, where $\oxygen = \{0, 1\}$ indicates the detection or non-detection of either $\oxygen$-A absorption (imaging mode) or $\ozone$ Chappuis band absorption (transit mode).

\subsubsection{Target prioritization}
Unlike in the previous example, we do not prioritize targets by size or insolation except that we assume all targets have previously been identified as EECs (perhaps with follow-up observations to confirm the presence of $\water$). This assumption is not trivial; imaging surveys cannot easily determine a planet's size, and the true range of planet sizes and insolations which permit habitability are not yet known. In reality, it is likely that an actual biosignature survey will probe some planets which are not habitable for reasons yet unknown to the observer, which will serve as a source of noise (i.e. by reducing $\flife$).

However, we do prioritize targets by age according to Figure \ref{fig:ex2_priority}, with observations of the youngest and oldest planets being preferred. This is not intended to counter any bias in the underlying sample, as there are no factors which bias the number of planets which can be characterized by our simulated surveys as a function of age. Rather, as we demonstrate in \citetalias{Bixel2020b}, a survey which prioritizes younger and older planets will be more sensitive to monotonic, age-dependent trends because of the larger contrast between those categories. While this prioritization strategy is optimal for studying the evolution of Earth-like planets, it must be balanced versus the survey's other goals. Notably, it de-prioritizes observations of modern Earth analogs, which may be the best planets to probe if the sole goal is to maximize the chance of detecting $\oxygen$.

\subsubsection{Time budget}
As discussed in Section \ref{sec:ex2_measurements}, we consider the detection of $\oxygen$-A absorption in imaging mode and $\ozone$ Chappuis band absorption in transit mode. The details of the exposure time calculations are shown in Table \ref{tab:exptimes}. Using PSG, we determine the exposure time required for the reference target to be $\tref = 2.6$ hr for imaging mode and $\tref$ = 74 hr for transit mode.

\subsection{Hypothesis and prior assumptions}
Once more, we take the role of an observer intrepreting the results of each simulated survey. Our hypothesis is that inhabited planets tend to evolve towards greater oxygen content over time, and can be stated in similar terms as Equation \ref{eqtn:foxygen}:
\begin{equation} \label{eqtn:h_age_oxygen}
    h(t_*) = \flife \left(1 - 0.5^{t_*/\thalf}\right)
\end{equation}

We adopt broad, log-uniform prior distributions for $\flife$ and $\thalf$, shown in Table \ref{tab:priors}, reflecting our significant prior uncertainty as to frequency and evolutionary timescales of inhabited planets.

\subsection{Correlation test}
In lieu of the Bayesian evidence test used in the previous example, we employ the Mann-Whitney test \citep{Mann1947} to determine whether $t_*$ correlates with the presence of oxygen, as we previously have done in \citetalias{Bixel2020b}. This model-independent test is more sensitive for detecting the correlation than the Bayesian evidence-based approach, especially in the limit of small sample sizes. However, it does not allow for the estimation of $\thalf$, for which we rely on MCMC sampling.

\begin{figure*}[ht]
\centering
\textbf{Example 2: Results for 15-meter imaging survey}

\textbf{How many exo-Earth candidates are probed for $\oxygen$?} \\
\includegraphics[width=0.49\textwidth]{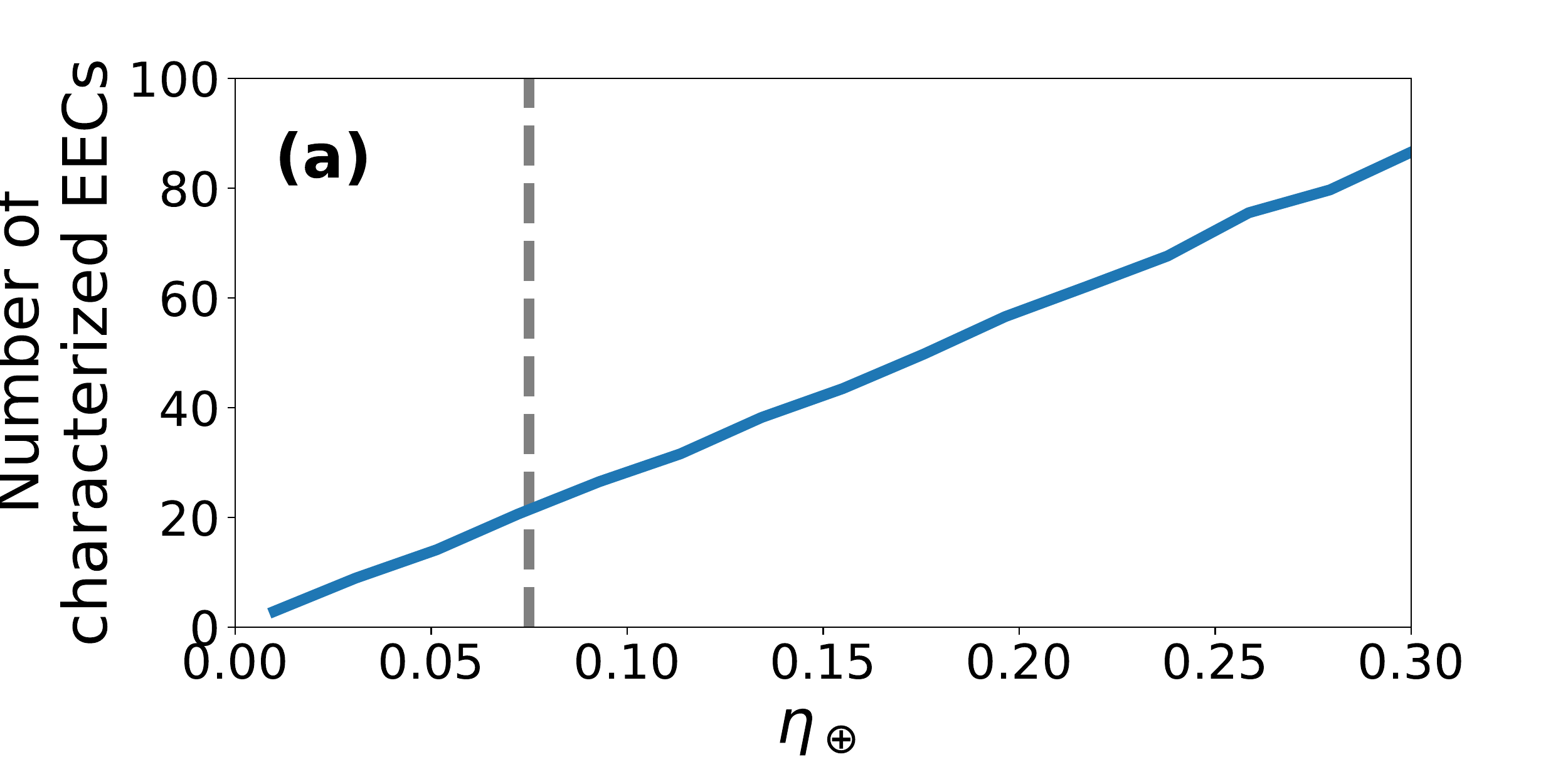}
\includegraphics[width=0.09\textwidth]{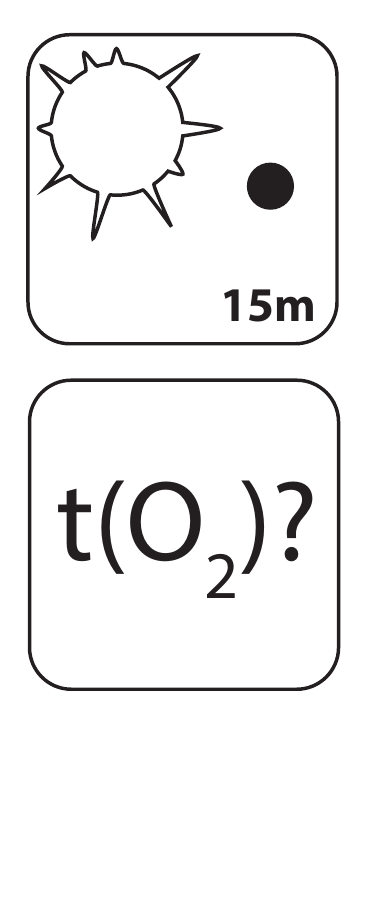}

\textbf{Could this survey detect the age-oxygen correlation?} \\
\includegraphics[width=0.49\textwidth]{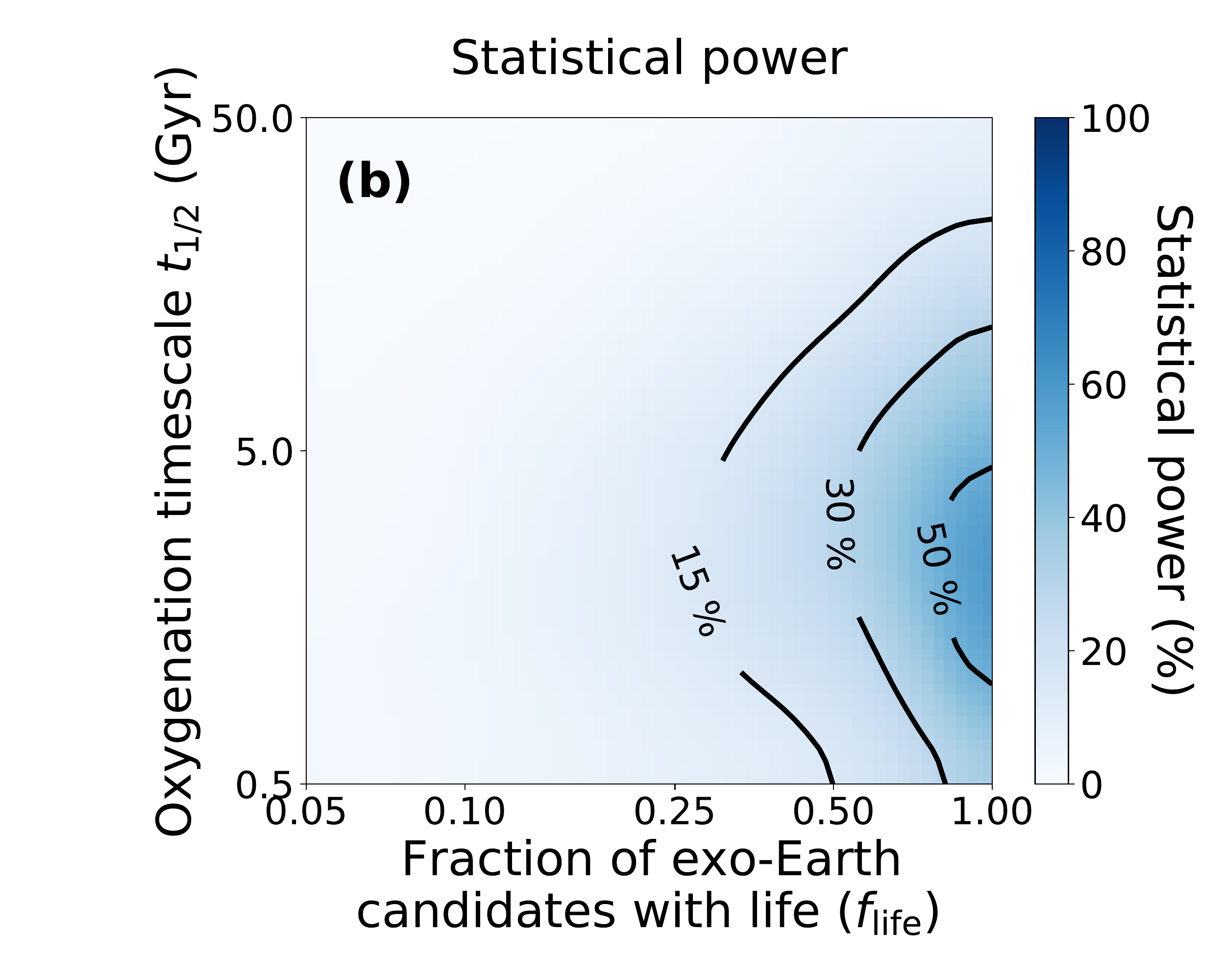}
\includegraphics[width=0.49\textwidth]{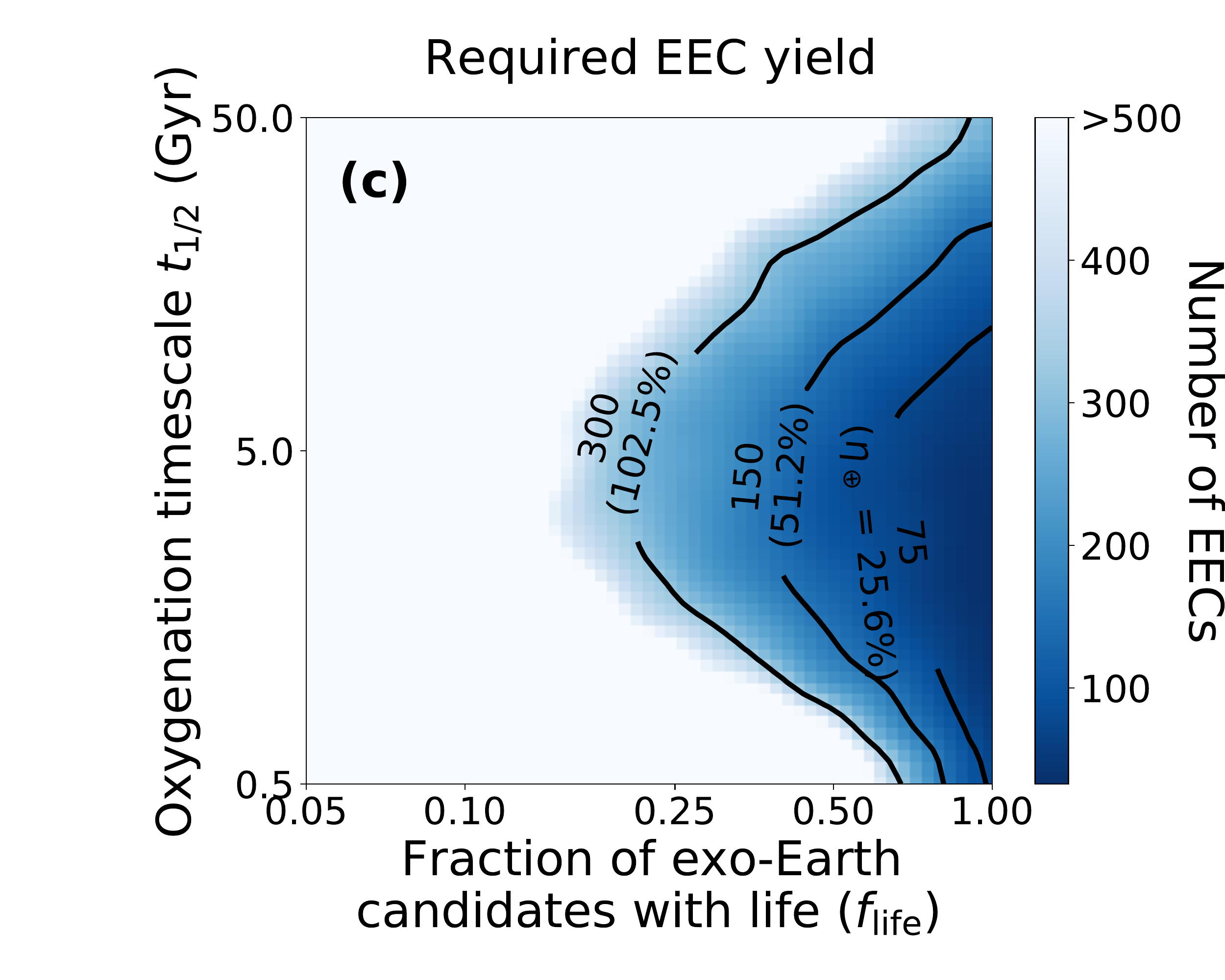}

\caption{Results for the imaging survey in Section \ref{sec:example2}. (a) The number of EECs observed versus $\ee$ (for G stars), assuming $\ttotal = 120$ d. For our baseline case, we set $\ee = 7.5\%$. (b) The statistical power to detect the age-oxygen correlation as a function of the astrophysical parameters in Equation \ref{eqtn:foxygen}. (c) The minimum number of EECs which must be characterized to achieve 80\% statistical power, with the corresponding values of $\ee$. \label{fig:example2_imaging}}
\end{figure*}

\begin{figure*}[p]
\centering
\textbf{Example 2: Results for 50-meter (equivalent area) transit survey}

\textbf{How many exo-Earth candidates are probed for $\ozone$?} \\
\includegraphics[width=0.49\textwidth]{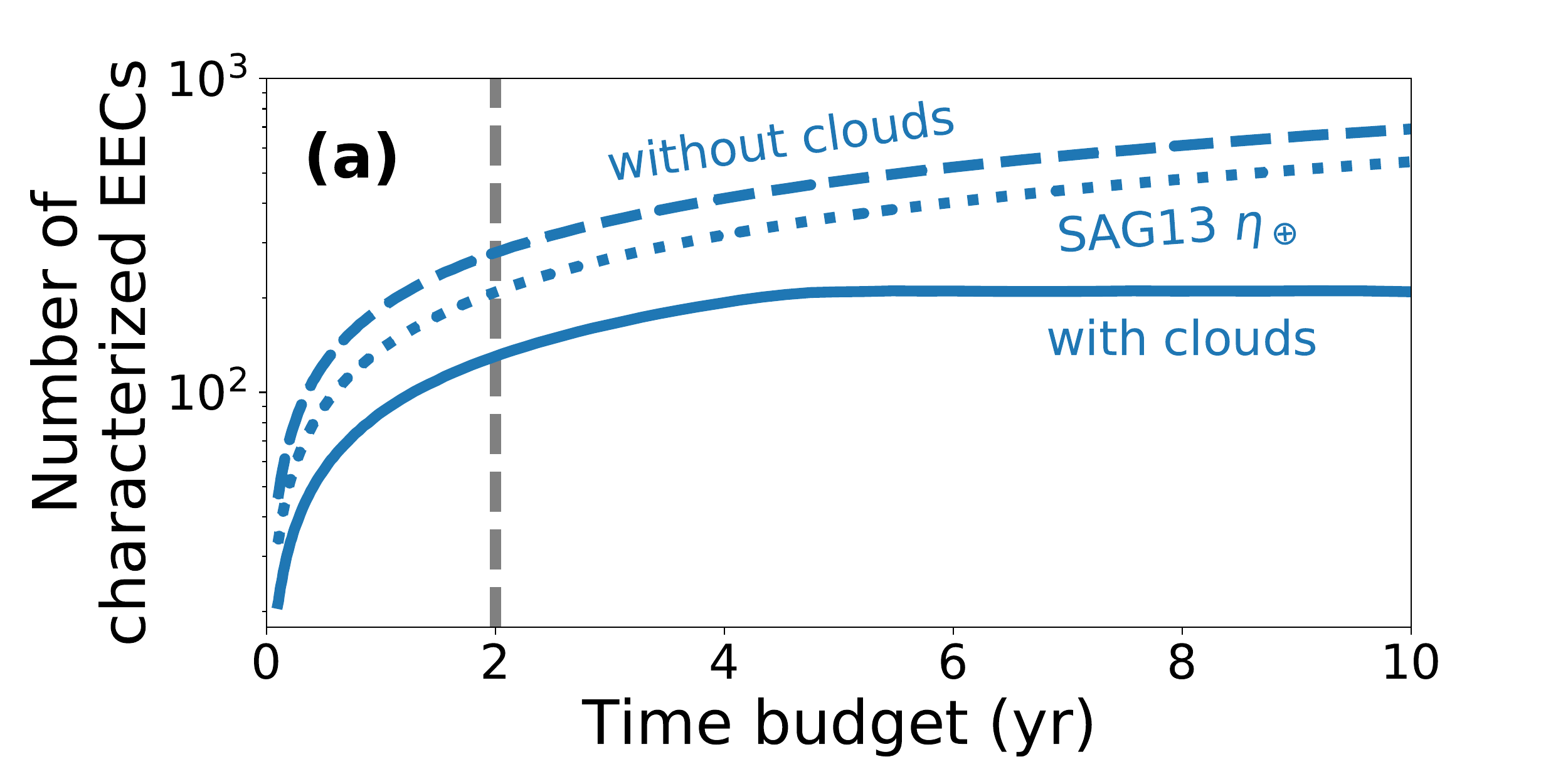}
\includegraphics[width=0.09\textwidth]{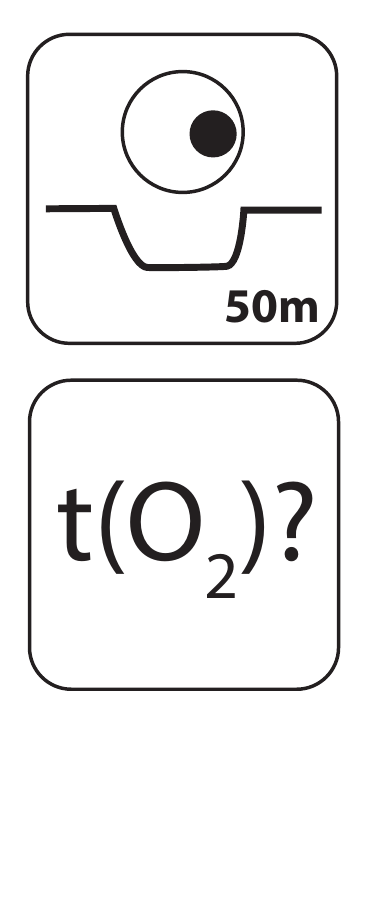}

\textbf{Could this survey detect the age-oxygen correlation?} \\
\includegraphics[width=0.49\textwidth]{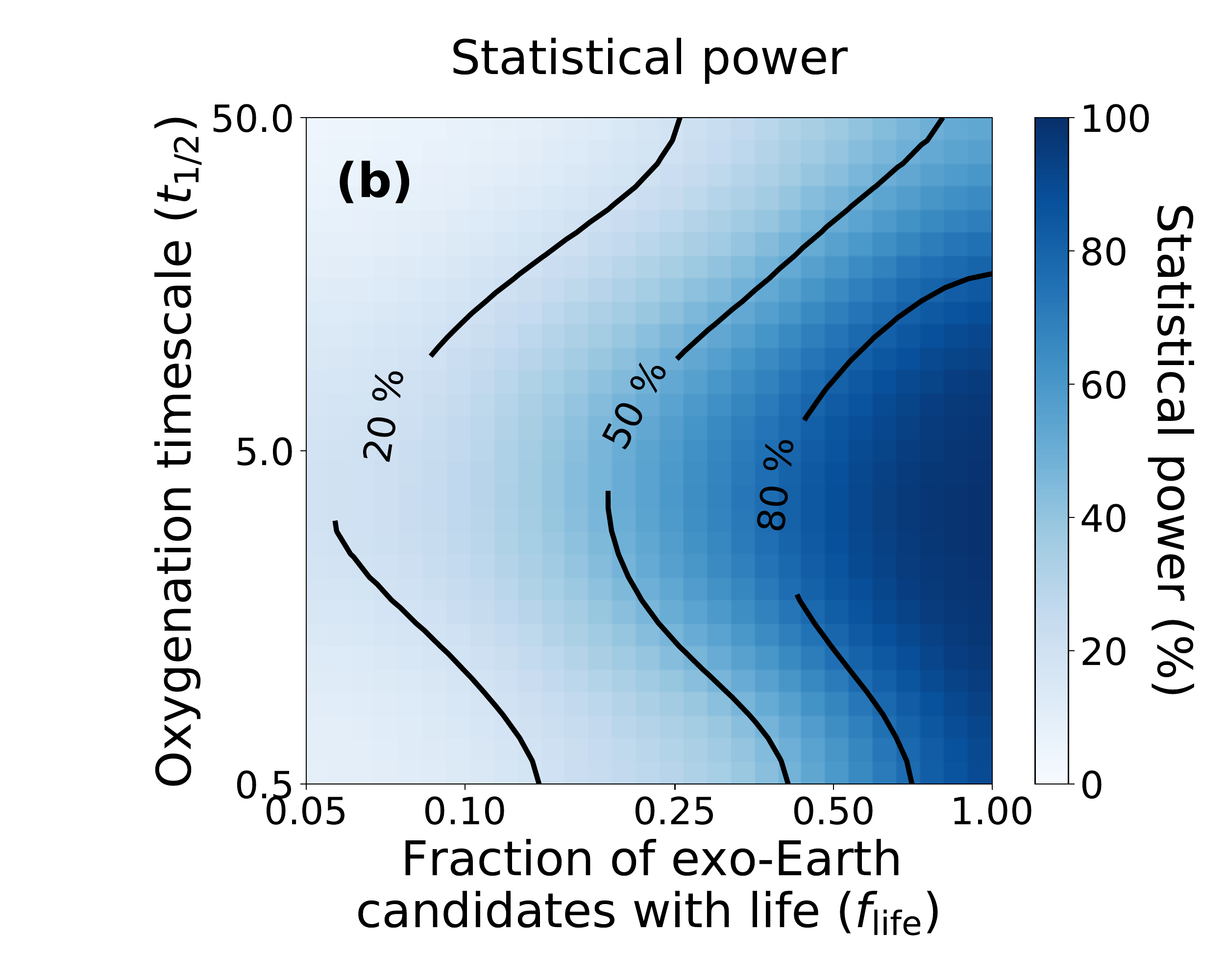}
\includegraphics[width=0.49\textwidth]{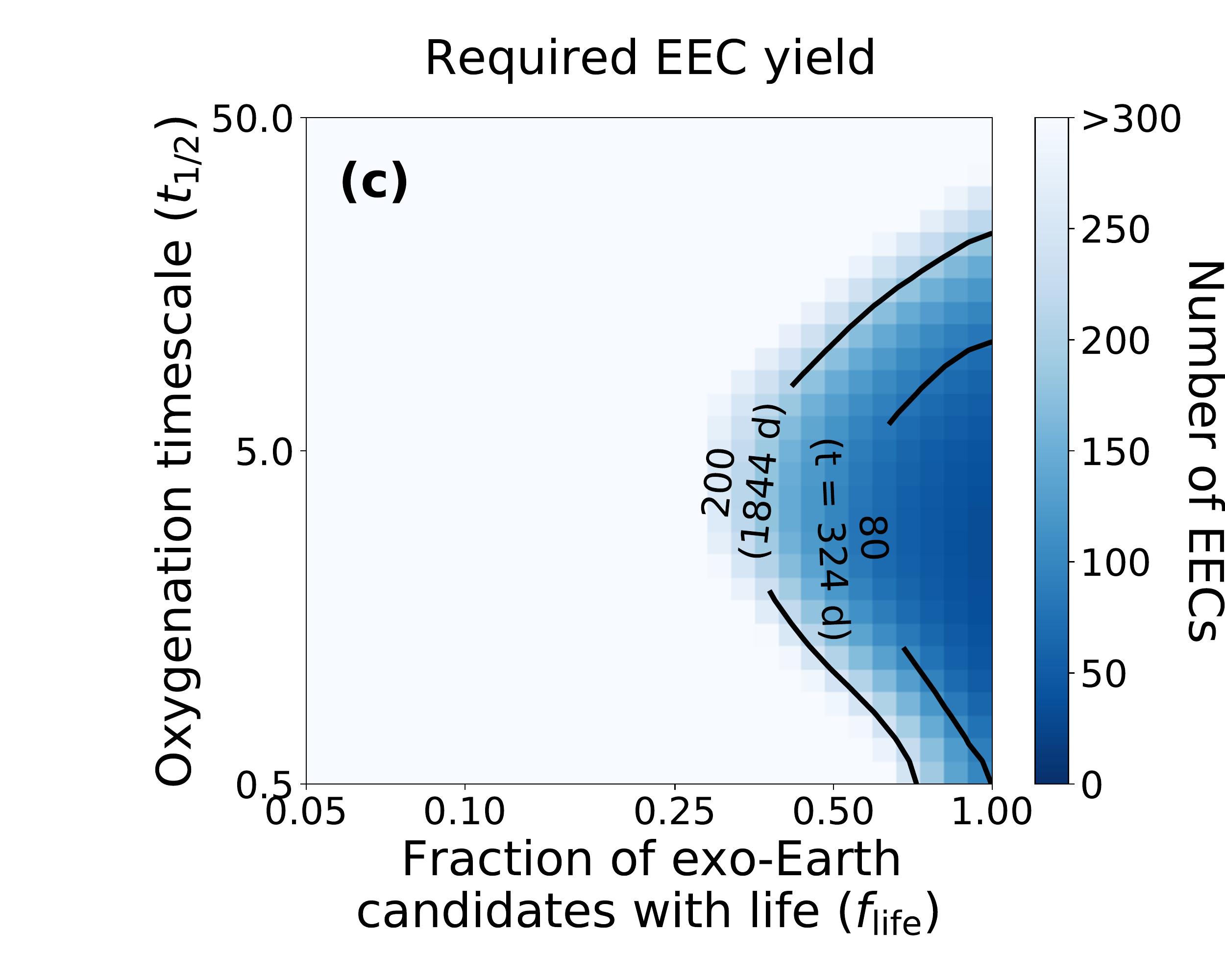}

\textbf{Could this survey determine the oxygenation timescale?} \\
\textbf{(results for six random realizations of the survey)}
\includegraphics[width=\textwidth]{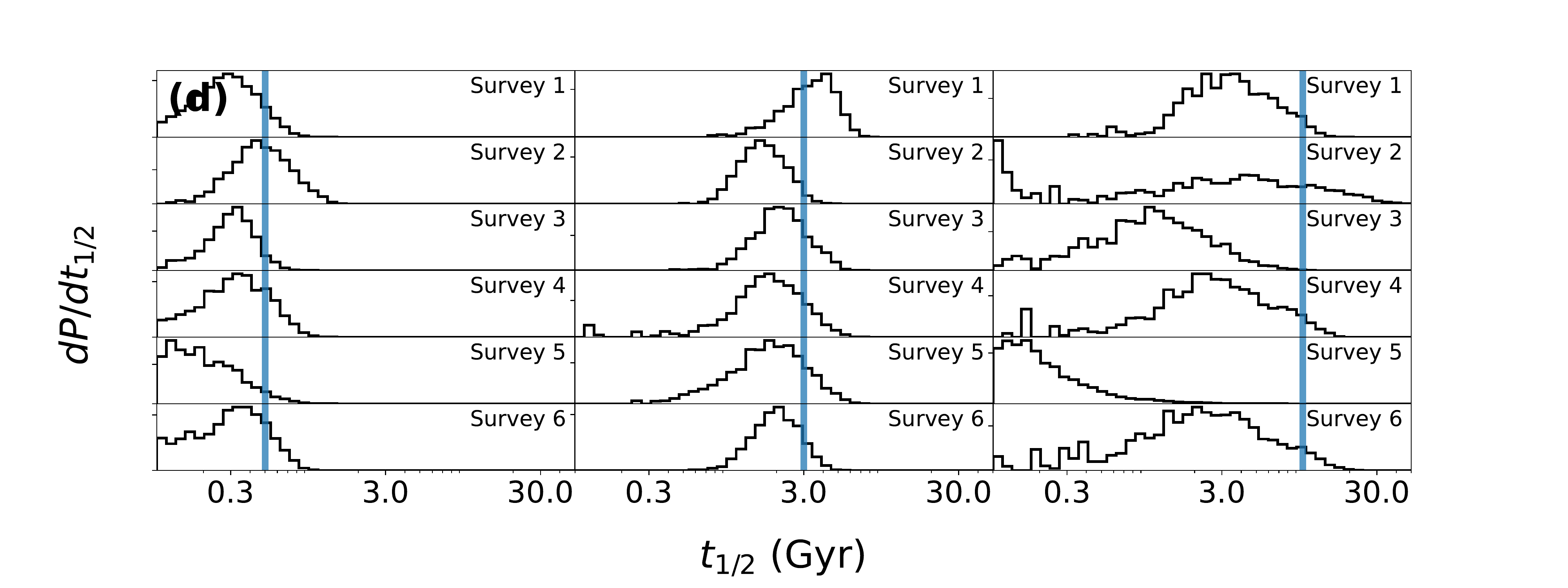}

\caption{Results for the transit survey in Section \ref{sec:example2}. (a) The number of EECs observed versus the observing time budget, assuming $\ee = 7.5\%$ for G stars and cloudy atmospheres (solid). 2--4$\times$ as many planets could be observed if clouds were neglected (dashed), or 1.5--3$\times$ as many with clouds if assuming the higher SAG13 estimate of $\ee = 24\%$ (dotted). As our baseline case, we set $\ttotal = 730$ d. (b) The statistical power to detect the age-oxygen correlation as a function of the astrophysical parameters in Equation \ref{eqtn:foxygen}. (c) The minimum number of EECs which must be characterized to achieve 80\% statistical power, with the necessary observing time budget $\ttotal$. (d) Distribution of possible values for the oxygenation timescale as measured by six random realizations of the survey under the optimistic assumption that 80\% of EECs are inhabited. Results are shown for fast (0.5 Gyr, left), Earth-like (3 Gyr, center), and slow (10 Gyr, right) evolutionary scenarios, with the truth values marked by a blue line. \label{fig:example2_transit}}
\end{figure*}

\subsection{Results}
We assess the statistical power of each survey to test the age-oxygen correlation hypothesis, using the Mann-Whitney test to determine whether a positive correlation can be detected in each simulated data set, and \texttt{emcee} to sample the posterior distributions of $\thalf$ and $\flife$. Our results are summarized in Figure \ref{fig:example2_imaging} for the imaging survey and Figure \ref{fig:example2_transit} for the transit survey.

\subsubsection{Imaging survey}
Assuming $\ee \approx 7.5\%$, it is unlikely (though not impossible) that a direct imaging survey will be able to detect the age-oxygen correlation with a sample of 15--20 EECs. This is generally consistent with our analysis in \citetalias{Bixel2020b}, which suggests a statistical power of 50\% for a sample size of $\sim 20$ EECs only if most are inhabited. In order for an imaging survey to be reliably capable of studying the oxygen evolution of Earth-like planets under optimistic circumstances, a sample size of $> 50$ EECs is necessary, requiring either $\ee \gtrsim 20\%$ or a smaller inner working angle than assumed here (3.5 $\lambda$/$D$).

\subsubsection{Transit survey}
By probing 100-150 EECs for ozone, the transit survey is able to detect the age-oxygen correlation with high statistical power assuming life to be somewhat common ($\flife \gtrsim 50\%$) and the typical oxygenation timescale to be 1--10 Gyr. If life is very common ($\flife \gtrsim 80\%$), high statistical power can be achieved even if the average oxygenation timescale is as short as $\sim 500$ Myr or as long as $\sim 20$ Gyr.

Under the case where life is very common, the transit survey could place meaningful constraints on the oxygenation timescale. As shown in Figure \ref{fig:example2_transit}, the survey can distinguish between scenarios where global oxygenation occurs very quickly ($\thalf \sim 0.5$ Gyr) or at a more Earth-like pace ($\sim 3$ Gyr), but it will be difficult to accurately measure the oxygenation timescale if it is much longer than Earth's ($\gtrsim 10$ Gyr), since no planets of that age exist in the sample. This is due in part to the high degeneracy between $\thalf$ and $\flife$ - that is, if only a few oxygenated planets are found, it may be because life is uncommon, or because life is common but global oxygenation is very slow and has not yet had time to occur on most inhabited worlds.

\subsection{Discussion}

\subsubsection{Detectability of oxygen through Earth's history} \label{sec:ozone}
In this section we consider all oxygenated planets to have the same $\oxygen$ and $\ozone$ abundance as modern Earth. However, during the Proterozoic era (approx. 2.2 -- 0.6 Gya), Earth had a partially oxygenated atmosphere with $p\oxygen < 1\%$ \citep{Lyons2014}. If other inhabited planets do evolve like Earth, this suggests that many of them may have 1--3 orders of magnitude less $\oxygen$ than modern Earth.

In our analysis, $\thalf$ is the typical timescale require for a planet to achieve a detectable amount of $\oxygen$ or $\ozone$. Even if Proterozoic Earth analogs are common and their oxygen is undetectable, our results should not be affected provided that they will eventually develop richly-oxygenated atmospheres like modern Earth's. In this case $\thalf$ corresponds to the end of the Proterozoic ($\sim 4$ Gyr for Earth). On the other hand, it may be that inhabited and oxygenated planets are common but very few of them ever evolve beyond $p\oxygen = 0.1 - 1\%$, in which case $\thalf$ corresponds to the end of the Archean ($\sim$ 2 Gyr for Earth). In this case, a survey aiming to detect the age-oxygen correlation would need to focus on a smaller number of targets with much deeper observations, and would likely need ultraviolet sensitivity to detect the deep $\ozone$ Hartley band absorption which would have been detectable throughout the Proterozoic \citep{Reinhard2017}. For transit spectroscopy, ultraviolet sensitivity will be difficult to achieve in a sample of predominantly M stars, so to detect $\ozone$ at Protorezoic-like levels will require the prioritization of G and K targets instead. A LUVOIR-like direct imaging survey targeting G and K dwarfs may be capable of detecting Proterozoic-like ozone levels for individual targets, but the sample size will still be too small unless both $\ee$ and $\flife$ are large ($\gtrsim 30\%$).

\subsubsection{Abiotic oxygen sources}
We only consider planets on which \Otwo\ is biologically produced - as it was in Earth's history - but others have considered scenarios through which an Earth-sized planet near or within the habitable zone could acquire detectable levels of oxygen through abiotic processes \citep[for a review, see][]{Meadows2018}. The oxygen in these models tends to initially derive from \water\ or CO$_2$ dominated atmospheres shortly after the planet's formation and can linger in the atmosphere long enough to serve as a potential ``false positive'' biosignature for next-generation observatories. In \citetalias{Bixel2020b}, by assuming the fraction of planets with abiotically produced oxygen to be independent of age, we show that these false positives will have a small impact on the detectability of the age-oxygen correlation provided that they are less common than Earth-like planets with biogenic \Otwo.

In reality, atmospheres with abiotically-produced oxygen will evolve over time. On Earth, oxygen is continually produced in large enough quantities to overcome its substantial geological sinks. On planets where oxygen is, e.g., a remnant of primordial ocean loss, it would be depleted over time. This suggests a statistical test to determine whether oxygen is a reliable biosignature: if the fraction of EECs with oxygen \emph{decreases} with age, this would suggest much of the oxygen to be of a primordial, abiotic origin.

Finally, it is plausible that both populations of oxygen-rich worlds exist in comparable numbers: one with abiotically-produced oxygen which diminishes over time, and another with biologically produced atmospheres which increases over time. Whether the Earth-like age-oxygen correlation could be detected would depend on the timescales of the two processes. For example, if most planets with abiotically produced oxygen lose it before 1 Gyr, and most planets with biogenic oxygen acquire it by 10 Gyr, then it should be possible to distinguish the two populations.

\section{Summary}

We have presented \codename, a simulation tool designed to gauge the potential of future observatories to test statistical hypotheses about the formation and evolution of planetary systems and habitable worlds. To achieve this, \codename\ leverages statistically realistic simulations of nearby planetary systems, a survey simulator designed to produce data sets representative of different observatory configurations and survey strategies, and a hypothesis testing module to assess the information content of the data. We demonstrated two applications of our code.

In the first example, we determined whether a future direct imaging (15-meter diameter) or transit spectroscopy (50-meter equivalent diameter) survey could empirically test the concept of a habitable zone as well as measure its location and width. With samples as small as 15--20 EECs, we found that both surveys will be capable of testing the habitable zone hypothesis if habitable planets are common ($\gtrsim 50\%$ of EECS), and that they can constrain the habitable zone's width well enough to rule out very wide (e.g., 1--10 AU) or narrow (e.g., 1--1.1 AU). A survey which can characterize 60--70 EECs for atmospheric water vapor can test the habitable zone hypothesis even if habitable planets are less common (20--40\% of EECs), but would be difficult to achieve with currently-envisioned direct imaging observatories. Our estimates suggest that this would be feasible for a large aperture transit survey, but the EEC sample size is sensitive to the impact of cloud cover (and other factors not considered here, such as stellar contamination \citep{Rackham2018}).

In the second example, we expanded upon the age-oxygen correlation proposed in \citetalias{Bixel2020b}, finding that future surveys which aim to study the oxygen evolution of Earth-like planets must expect to characterize at least $\sim 50$ EECs by detecting the presence of modern Earth-like $\oxygen$ or $\ozone$ absorption. With a sample size of 100-150 EECs -- if most of them are inhabited -- a survey could begin to constrain the evolutionary timescale with meaningful precision, and could determine whether the oxygenation of Earth-like planets proceeds at an Earth-like pace (2--5 Gyr timescale) or much faster ($\sim 0.5$ Gyr). The ability to detect far-UV $\ozone$ absorption will be beneficial if Proterozoic Earth analogs are common, but may not be necessary provided they eventually evolve to a modern Earth-like state.

The statistical power of either survey to test these hypotheses depends critically on the number of EECs detected, but recent evidence suggests that existing estimates of $\ee$ are too high \citep{Pascucci2019, Neil2020}. Assuming $\ee = 7.5\%$ for Sun-like stars, we found that an ambitious 15-meter mirror diameter imaging survey would likely detect 15--20 EECs. Such a survey may have high statistical power for studies of terrestrial planets in general (including those outside the habitable zone), but will only be able to test the habitable zone concept if most EECs are habitable, or if tracers of habitability other than $\water$ absorption are considered. Unless $\ee > 20\%$, an imaging survey will probably not be able to study the oxygen evolution of truly Earth-like planets, though it might still offer constraints on how common such planets are \citep{Checlair2020}.

In this paper we discussed the statistical power to test hypotheses as a function of sample size given a single measurement for each target. \codename\ can also combine multiple measurements for each planet which trace the same underlying physical conditions (such as habitability), allowing surveys to achieve greater statistical sensitivity with limited sample sizes. For example, by incorporating measurements of planetary brightness and color in addition to $\water$ absorption, imaging surveys may be able to test the habitable zone concept with smaller sample sizes - provided a hypothesis exists for how these properties should vary with orbital separation \citep[e.g.,][]{Checlair2019}. Similarly, if clouds make the detection of $\water$ difficult for a transit survey, then stratospheric $\ozone$ may offer an alternative tracer of planetary habitability (provided $\oxygen$ is predominantly produced by life).

With \codename, we aim to enable future space-based exoplanet surveys to test hypotheses including and beyond the examples explored here, and to emphasize the importance of population-level studies for next-generation exoplanet surveys. While target-by-target analyses of the closest planets will be valuable, population-level studies will reveal fundamental truths about the laws governing non-habitable, habitable, and inhabited worlds.

\acknowledgements
We thank Chris Stark for providing a possible realization of the LUVOIR-A target list and Tad Komacek for providing the GCM model outputs used in this work. A.B. acknowledges support from the NASA Earth and Space Science Fellowship Program under grant No. 80NSS\-C17K0470. The results reported herein benefited from collaborations and/or information exchange within NASA's Nexus for Exoplanet System Science (NExSS) research coordination network sponsored by NASA's Science Mission Directorate. This research has made use of NASA's Astrophysics Data System.

\software{\texttt{dynesty} \citep{Speagle2020}, \texttt{emcee} \citep{Foreman-Mackey2013}, Matplotlib \citep{Hunter2007}, NumPy \citep{Oliphant2006}, PSG \citep{Villanueva2018}, SciPy \citep{Virtanen2020}}

\bibliography{references}{}
\bibliographystyle{aasjournal}

\appendix

\section{List of symbols} \label{sec:appendix1}

\startlongtable
\begin{deluxetable}{lp{4in}}
\tablecaption{A list of common abbreviations and symbols used in this paper.}
\tablehead{Symbol & Description}
\startdata
\textbf{Abbreviations} & \\
EEC & ``exo-Earth candidate'' (or ``potentially habitable planet''); planets in the radius range $0.8(S/S_\oplus)^{0.25} < R_p < 1.4\, R_\oplus$ \\
LUVOIR & Large UV/Optical/Infrared Surveyor \citep{LUVOIR2019} \\
SAG13 & NASA's Exoplanet Program Analysis Group Science Analysis Group 13 \\
PSG & NASA/GSFC Planetary Spectrum Generator \citep{Villanueva2018} \\
IWA, OWA & Inner, outer working angles of a coronagraphic instrument \\
MCMC & Markov Chain Monte Carlo \\
& \\ \textbf{Stellar properties} & \\
$d$                 & Distance to star \\
$M_*, R_*, L_*$     & Mass, radius, and luminosity \\
$T_*$               & Effective temperature \\
$t_*$               & Age of star and planetary system \\
$\ainner$, $\aouter$ & Inner and outer edge of the star's habitable zone \\
& \\ \textbf{Planet properties} & \\
$M_p, R_p, g_p$     & Mass, radius, and surface gravity \\
$h$                 & Atmospheric scale height  \\
$P$                 & Orbital period \\
$a$                 & Semi-major axis \\
$\aeff$             & Semi-major axis scaled by the stellar luminosity, $\aeff = a (L_*/L_\odot)^{-0.5}$ \\
$\cos(i)$            & (Cosine of) orbital inclination \\
$b$                 & Transit impact parameter, assuming a circular orbit \\
$\delta$            & Planet transit depth, $\delta = (R_p/R_*)^2$ \\
$\Delta \delta$     & Approximate transit depth induced by planet's atmosphere, $\Delta \delta \sim 2(h/R_p)$ \\
$\zeta$             & Planet-to-star contrast ratio \\
$\Reff$             & Estimated planet radius assuming Earth-like reflectivity (direct imaging only), $\Reff/R_\oplus = (\zeta/\zeta_\oplus)^{0.5} (a/1\,\text{AU})$ \\
& \\ \textbf{Simulated survey} & \\
$D_\text{tel}$      & Telescope diameter or effective diameter (based on total light-collecting area) \\
$\leff$             & Effective wavelength of a spectroscopic measurement \\
$\Rref$, $\Tref$    & Radius and effective temperature of the reference star; ($\Rref, \Tref$) = (5777 K, $1\,R_\odot$) for the imaging survey, (3000 K, $0.15\,R_\odot$) for the transit survey \\
$t_i$               & Amount of time required to characterize the $i$-th planet in a sample \\
$\tref$             & Amount of time required to characterize an Earth twin orbiting the reference star with $\aeff = 1$ AU \\
$\ttotal$           & Time budget allocated to characterizing planets for a specific spectral feature (may overlap with observations at other wavelengths) \\
$\zeta_\oplus$      & Contrast ratio of the Earth with respect to the Sun, $\zeta_\oplus \approx 10^{-10}$ \\
$p_i$, $w_i$        & Observing priority and relative weight assigned to each planet, where $p_i = w_i/t_i$ \\
& \\ \textbf{Hypothesis testing} & \\
$x$, $y$            & Independent and dependent variables in the simulated data sets \\
$h(\vtheta, x)$     & Alternative hypothesis describing the relationship between $x$ and $y$, to be compared to the null hypothesis \\
$\vtheta$           & Set of parameters which define $h$ \\
$\mathcal{L}(y|\vtheta)$       & Likelihood function, described by Equation \ref{eqtn:likelihood} or \ref{eqtn:likelihood2} \\
$\Pi(\vtheta)$      & Prior probability distribution of $\vtheta$, described for each example in Table \ref{tab:priors} \\
$\mathcal{Z}$       & Bayesian evidence in favor of the null or alternative hypothesis, computed by nested sampling \\
& \\ \textbf{Habitable zone hypothesis} & \\
$\ainner$, $\aouter$ & Inner and outer edges of the habitable zone in $\aeff$ space (i.e. for a Sun-like star) \\
$\fwaterhab$         & Fraction of EECs with atmospheric water vapor (assumed habitable) \\
$\fwaternonhab$      & Fraction of non-EECs with atmospheric water vapor \\
& \\ \textbf{Age-oxygen correlation} & \\
$\flife$             & Fraction of EECs inhabited by life (regardless of $\oxygen$ content) \\
$\thalf$             & Oxygenation timescale; the time required for 50\% of inhabited planets to undergo global oxygenation \\
\enddata
\end{deluxetable}
\quad

\end{document}